\documentclass[journal, a4paper]{IEEEtran}

\usepackage{cite}
\usepackage{amsmath,amssymb,amsfonts}
\usepackage{tikz}
\usetikzlibrary{arrows,automata}
\usepackage{caption}
\usepackage{graphicx}
\usepackage{textcomp}
\usepackage{xcolor}
\usepackage{algorithm}
\usepackage{algpseudocode}
\usepackage{tcolorbox}
\usepackage{multirow}
\usepackage{cuted} 

\usepackage[margin=.4in]{geometry}
\begin{document}
	\IEEEpubid{\makebox[\columnwidth]{\textit{Submitted to Digital Signal Processing, Elsevier}} \hspace{\columnsep}\makebox[\columnwidth]{ }}

	\title{Multifold Confidence Intervals in Collaborative Mean Estimation (ColME) Using Sample Statistics}
	
	\author{Nikola~Stankovi\' c
		\thanks{N. Stankovi\' c is with the University of Montenegro, Podgorica, Montenegro (email: nikola.stankovic1@edu.ucg.ac.me).}%
		\thanks{The author would like to thank Prof. Emilio Leonardi and Dr. Franco Galante from Politecnico di Torino, Italy, for introducing him to this topic and for their valuable comments during his Master's thesis work.}%
	}

	\maketitle
	
	
	\begin{strip}
		
	\vspace{-65pt} 
	
		\begin{abstract} 
			\textnormal{The rapid growth of personal digital devices and the Internet of Things (IoT) has intensified the demand for collaborative and decentralized learning frameworks. Since these devices continuously generate sensitive and high-dimensional data, centralized transmission is often impractical, while purely local learning suffers from slow convergence as data accumulate gradually. Collaborative approaches can alleviate these issues by allowing agents to use information from one another to improve estimation, yet the challenge lies in the heterogeneity of data distributions. Each agent typically faces a personalized learning problem, and collaboration is only beneficial among agents whose data are generated from the same distributions, meaning they belong to the same similarity class. This paper studies the problem of personalized online mean estimation in such heterogeneous environments, where each agent observes data from its own $\sigma$-sub-Gaussian distribution. To address the challenge of heterogeneity, collaborative algorithms enable agents to identify similarity classes in real time and exploit information from agents belonging to the similarity class within the same class to accelerate convergence and improve accuracy. The work builds on existing approaches: the collaborative mean estimation (colME), which refines estimates through agent interaction using confidence intervals, and its graph-based extensions (C-colME and B-colME), which improve scalability and robustness in distributed settings.  Since the variance estimation plays a crucial role in the above mentioned algorithms, a method for accurate, local and real-time estimation of variance is proposed in this paper. A mean-independent and online variance estimator is introduced. Estimation of sample kurtosis is also incorporated without explicit reliance on the agents’ means, enabling the design of refined confidence intervals that account for differences not only in means and variances but also in higher-order characteristics (distribution) of the data. We derive the confidence interval estimators for the sample standard deviation and sample kurtosis. These results are combined with sample mean estimation methods to design a unified procedure for constructing multifold confidence intervals based jointly on the sample mean, sample variance, and sample kurtosis. This framework enables collaborative estimation in challenging scenarios, such as when classes share similar means but differ in variances, or when both means and variances are alike while the underlying distributions diverge in higher-order characteristics, not considered up to now. The oracle scenario is used as a benchmark for the accuracy and convergence rate of the proposed approach. Theory is illustrated on numerical examples. 
	}
		\end{abstract}
		
		\vspace{10pt} 
		\hrule        
		\vspace{10pt}
	\end{strip}
	\IEEEpubidadjcol

\section{Introduction}
The advances and widespread applications of personal digital devices, coupled with progress in computing and the Internet of Things (IoT), have amplified the need for decentralized and collaborative computational paradigms. These devices are inherently data-centric, generating information that is often sensitive or voluminous, rendering centralized transmission impractical or undesirable. Although isolated local processing remains an option, learning in such a solitary manner can suffer from slow convergence, particularly in online settings where data accumulates gradually.

Over the past few years, the field of Federated Learning (FL) has witnessed extensive research into collaborative estimation and learning problems involving multiple agents with local datasets \cite{kairouz2021advances,tan2022towards}. While traditional FL approaches focus on learning a single global model applicable to all participating agents, the inherent statistical heterogeneity across client data has spurred significant interest in personalized FL techniques. These aim to develop models tailored to the specific data distributions of individual clients, with the objective of improving estimation accuracy and convergence behavior \cite{ghosh2020efficient, fallah2020personalized,sattler2021clustered,li2021ditto,sattler2021clustered,marfoq2021federated,ding2022collaborative}.

A common strategy in personalized FL involves grouping clients into clusters and subsequently training a specific model for each cluster \cite{ghosh2020efficient, sattler2021clustered,ding2022collaborative}. The ideal clustering would group clients possessing similar local optimal models, enabling effective collaboration and faster convergence. However, as these optimal models are typically unknown a priori, the processes of model learning and cluster identification become intrinsically linked. Several studies have proposed using empirical measures of similarity, such as the Euclidean distance between local models or the cosine similarity of their updates, as a practical workaround \cite{ghosh2020efficient,sattler2021clustered}. Others assume some form of prior knowledge about the distances between clients' data distributions \cite{ding2022collaborative,even2022sample}. Nevertheless, accurately estimating these distances, particularly in online FL settings where clients may have limited local data, remains a significant challenge, as highlighted in \cite{even2022sample}. As a consequence, incorrect similarity estimation may substantially slow convergence by either preventing beneficial collaboration or promoting the exchange of misleading information.

In the online learning setting, collaborative learning has been predominantly explored in the context of multiarmed bandits (MAB). However, most existing approaches consider a single MAB instance solved collaboratively by multiple agents, where collaboration serves to accelerate learning toward a shared objective. Collaboration is often achieved through broadcast messages \cite{hillel2013distributed,tao2019collaborative}, via a central server \cite{wang2020aoptimal}, or through local message exchanges over a network graph \cite{sankararaman2019social,martinez2019decentralized, wang2020distributed,landgren2021distributed,madhushani2021one}. In all these cases, agents share a common learning goal, and collaboration primarily reduces uncertainty rather than addressing heterogeneity.

More recently, research has begun to address collaborative MAB settings where the arm means vary across agents. Extensions of \cite{boursier2019sic} examine scenarios with heterogeneous rewards under collision models, while other works consider federated or partially personalized objectives \cite{shi2021federated,karpov2022collaborative,reda2022near}. A key distinction from our work is that these approaches do not require the online identification of relationships between local distributions in order to solve the respective problems, as similarity structure is either assumed or not central to convergence.

The problem of mean estimation in a federated setting has been recognized as a fundamental building block \cite{dorner2024incentivizing,tsoy2024provable,grimberg2021optimal} with practical significance in domains such as smart agriculture, grid management, and healthcare \cite{adi2020machine}. In this context, Collaborative Mean Estimation (ColME) was introduced in \cite{asadi2022collaborative} as an online, decentralized framework where agents exchange information with peers to improve their mean estimates. However, ColME suffers from scalability limitations, and its convergence behavior is highly sensitive to heterogeneity across agents and to how collaboration is structured.
The work in \cite{galante2025scalable} builds upon collaborative mean estimation by developing scalable algorithms for multidimensional data and broader distribution classes, where agents self-organize based on perceived similarity. Graphs  \cite{ortega2018graph, leus2023graph,hammond2018spectral,jestrovic2017fast,stankovic2019graph,stankovic2020vertex} are used as the interaction domain with message-passing and consensus-based variants \cite{galante2025scalable}. Convergence speed is strongly influenced by how quickly agents can identify useful collaborators and how effectively information is aggregated over time.

A key challenge in collaborative mean estimation therefore lies in identifying similarity structure online in a way that enables effective collaboration while accelerating convergence. While collaboration among sufficiently similar agents can significantly reduce estimation error, incorrect or overly conservative collaboration strategies can delay learning. This challenge is particularly pronounced in online settings, where data arrive sequentially and decisions about collaboration must be made under uncertainty. The confidence intervals, \cite{lerga2009nonlinear, lerga2018adaptive, katkovnik1998instantaneous, khan2022convolutional, sadeghi2020window,das2020detection}, are used as a key tool for the similarity classes determination  \cite{asadi2022collaborative}.

This paper addresses the challenges of the confidence interval determination by studying mechanisms of the online, local, and collaborative estimation of the variance and other data statistics that enable the efficient application of the confidence intervals and adaptive graph-based collaboration. 

In this paper, a  method for the estimation of sample variance is proposed in the collaborative setting of mean estimation, without using the agents'  mean value as the final goal. Estimation of the variance is done locally. Collaboration with neighboring agents can be used to improve the accuracy of the estimate.  The cases where the variance is the same for the whole system and when it differs for different classes are considered.  In addition, the sample kurtosis estimation is implemented without using the agents' means.
The standard deviation of the estimated sample standard deviation and kurtosis is derived. This result, along with sample estimation methods, is used to propose a new combined method when confidence intervals are used based on the sample mean, sample standard deviation, and sample kurtosis. By this approach, we were able to consider and use collaborative estimation in very complex cases, not considered in the literature up to now, such as when the classes have similar means but different standard deviations, or the cases when classes have similar means and standard deviations, but the distribution of data differs, resulting in different kurtosis. Experiments are presented where all three sample estimators must be used for a successful similarity class separation.  To maintain the confidence intervals (thereby avoiding false pruning) while still achieving fast convergence, the weights of data points are introduced for data nearing the ends of these intervals. To achieve this goal, the weighted graphs are used, employing a Gaussian kernel as the graph weight.  The theory is numerically verified.

The paper is organized as follows. Section II introduces the problem formulation and collaborative mean estimation framework. Section III presents the proposed methods for sample variance and kurtosis estimation. Section IV gives a review of the collaborative methods for mean estimation. Multifold confidence intervals are introduced in Section V. Multiclass examples are given in Sections VI and VII. Section VIII concludes the paper.

\section{Problem Definition}

Consider $A=N$ agents that receive data and can perform some processing, along with communication with other agents. The key task considered here is to identify the set of agents with similar or the same data distribution. After this set is detected, collaborative processing, by exchanging the data, can significantly improve estimation convergence and accuracy.

We assume that each agent $a \in \mathcal{A}=\{1,2,,\dots, A\}$, receives a random sample $x_a(t) \in \mathbb{R}^1$, drawn from a distribution $D_a$ with expected mean $\mu_a=\mathbb{E}\{x_a(t)\}$.
The vector form of the signal is:
\begin{equation}
	\mathbf{x}(t)=\begin{bmatrix}
		x_{1}(t),
		x_{2}(t),
		\cdots, 
		x_{A}(t)
	\end{bmatrix}^T.
\end{equation} 

\noindent {\bf Similarity Classes:} Agents that receive random samples $x_a(t) \in \mathbb{R}^1$, drawn from the same distribution $D_a$,  with the same expected mean $\mu_a$  form a similarity class. The similarity class of an agent $a$ is denoted by $\mathcal{C}_a$.

We will also assume that among all agents there are $C$ similarity classes. Two agents $a$ and $a'$ belong to the same similarity class if $\Delta_{aa'}=|\mu_a -\mu_{a'}|=0$ holds, that is, if the agents share the same mean value.

In collaborative mean estimation (colME), the goal of each agent $a \in \mathcal{A}$  is to estimate its mean as fast as possible, using its own data and the data of some collaborative agents belonging to the same similarity class $\mathcal{C}_a$. 

\begin{figure}[htbp]
	\centering
	\includegraphics[scale=0.75, trim={0 0 5.6cm 0}, clip]{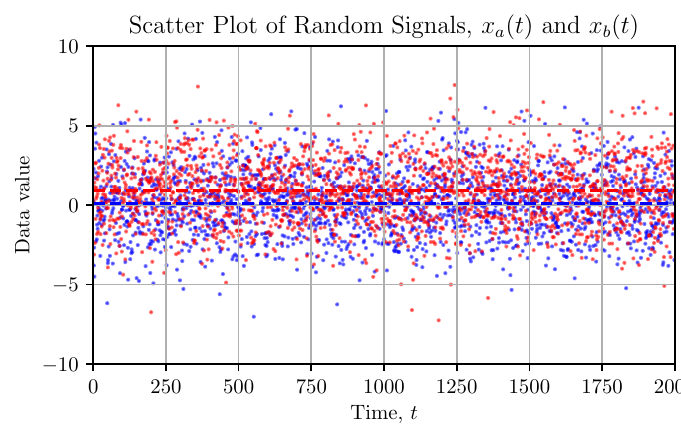}
	
	\caption{ Illustration of Gaussian data, $x_a(t)$, $x_b(t)$, of two agents, $a$ and $b$, from different similarity classes (shown in red and blue dots) for $t \in [0,2000]$ with $\mu_a=0.1$, $\mu_b=0.9$, $\Delta_{ab}=0.8$, and $\sigma=2$.}
	\label{Fig_DILL}
\end{figure}

\subsection{Local Means} 

The mean of one agent, after $t$ samples are acquired (the sampling step is equal to 1, that is, $t=1,2,3, \dots,T$), is called \textit{local mean}. It is defined as:
\begin{equation}
	\bar{x}_{a,a}(t) = \frac{1}{t}\sum_{\tau=1}^{t}x_a(\tau) = \frac{1}{t}x_a(t)+\frac{t-1}{t}\bar{x}_{a,a}(t-1).
\end{equation}
This empirical local mean slowly approaches the true mean $\mu_a$. Convergence can be significantly improved if we can determine the similarity class, $\mathcal{C}_a$, of the agent $a$, or at least a subset of agents from this class. Then we can use more data to improve the mean estimate convergence, by averaging the data not only over the time but over the set of data with the same mean (collaborative set). 

\begin{figure}[hptb]
	\centering
	\includegraphics[scale=0.75, trim={0 0 5.9cm 0}, clip]{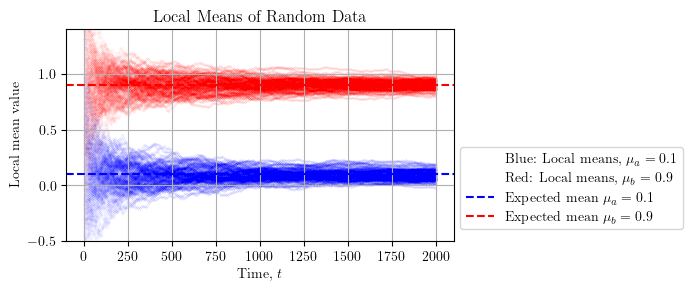}
	\caption{ Illustration of local means, $\bar{x}_{aa}(t)$, of data $x_a(t)$ with 2 similarity classes and $N=200$ agents (shown in red and blue), with $\mu_a=0.1$, $\mu_b=0.9$, $\Delta_{ab}=|\mu_a-\mu_b|=0.8$, for $t \in [0,2000]$. Data $x_a(t)$ are Gaussian distributed with the standard deviation $\sigma=2$.}
	\label{Fig_LM_ILL}
\end{figure}

The mean values for all agents, $a=1,2,\dots,A=N$, can be written in vector form as:
\begin{equation}
	\mathbf{X}(t)=\begin{bmatrix}
		\bar{x}_{1,1}(t) \\
		\bar{x}_{2,2}(t) \\
		\vdots \\
		\bar{x}_{A,A}(t) \\
	\end{bmatrix} \text{ with } \mathbf{X}(t)=\frac{1}{t}\mathbf{x}(t)+\frac{t-1}{t}\mathbf{X}(t-1).
\end{equation} 
Illustration of many realizations of local means for data with two similarity classes, $\mu_a=0.1$ and $\mu_b=0.9$, are shown in Fig. \ref{Fig_LM_ILL}. The local means for $N/2=100$ agents with the mean value $\mu_a=0.1$ are shown in red lines, while the local means corresponding to the same number of agents with $\mu_a=0.9$ are given in red. We can see that the realizations of local means can overlap for small $t$ and then very slowly converge to their true mean values as time $t$ progresses. 

In some approaches intead of local mean, 	$\bar{x}_{a,a}(t)$, a closely relation notion of local sum can be also used. It is defined as:
\begin{equation}
	m_{a,a}(t) =\sum_{\tau=1}^{t}x_a(\tau) = x_a(t)+ m_{a,a}(t-1)=	t\bar{x}_{a,a}(t) .
\end{equation}

\noindent \textbf{Oracle solution} for the estimated mean would be obtained if all similarity classes were known in advance. Then, at each instant, the estimated oracle mean would be obtained by averaging all local means, $\bar{x}_{a,a}(t)$, over all agents within the same similarity class, $a \in \mathcal{C}_a$.  This oracle (the best possible) value will be used to test the accuracy of the proposed solutions in numerical examples. 

\subsection{Confidence Intervals of Local Means}

For the estimation of a set of agents that belong to the same similarity class, as the considered agent $a$,  confidence intervals will be used. 

At each instant, $t$, after the agents receive new samples $x_a(t)$, that is a new vector $\mathbf{x}(t)$ is formed, they update their local means.
Then, each agent compares its local mean with the local means of its neighbors (the definition of neighborhood will be specified by an algorithm) in order to try to conclude if the considered neighbor is within the same similarity class. 

In general, the confidence intervals for a random variable  (in this case, the local sample mean, $\bar{x}_{a,a}(t) = \frac{1}{t}\sum_{\tau=1}^{t}x_a(\tau)$)  are defined as 
\begin{equation}\mathbb{I}_a(t)=[\bar{x}_{a,a}(t)-\beta_{\delta}(t),  \bar{x}_{a,a}(t)+\beta_{\delta}(t)],\end{equation} 
where $\beta_{\delta}(t)$ specifies the confidence interval width that will be discussed next.

Consider a random variable $\bar{x}_{a,a}(t)$ with a symmetric probability density function with respect to its mean $\mu_a$. The confidence intervals are defined based on the probability that a value of the random variable $\bar{x}_{a,a}(t)$ differs from its mean value $\mu_a$ is lower than a given small probability $\delta$. 

Based on 
\begin{equation}\mathbb{P}\Big(|\bar{x}_{a,a}(t)-\mu_a| > \sigma B_{\delta}(t)\}\Big)<\delta,\end{equation}
where $\sigma$ is the standard deviation of $x_a(t)$, we can state that the random variable $\bar{x}_{a,a}(t)$ satisfies the inequality 
\begin{equation}\mu_a -  \sigma B_{\delta}(t)\  \le \bar{x}_{a,a}(t) \le   \mu_a + \sigma B_{\delta}(t),\end{equation}
with probability greater than $1-\delta$. 

For the case where  $\lim_{t \to \infty }B_{\delta}(t)=0$, 
we can state that the mean, $\mu_a$, estimation based on sample mean, $\bar{x}_{a,a}(t) = \frac{1}{t}\sum_{\tau=1}^{t}x_a(\tau)$, is a Probably Approximately Correct (PAC) problem: We can make the estimation error smaller than an arbitrary small $\varepsilon$ with probability higher than $1-\delta$ for any small $\delta$, that is for any small $(\varepsilon, \delta)$ there exits $\tau_a$ such that for any $t>\tau_a$ we have 
\begin{equation}\mathbb{P}\Big(|\bar{x}_{a,a}(t)-\mu_a| <  \varepsilon \Big)>1-\delta.\end{equation}
This fact will be used later to define two crucial properties of the confidence intervals that will be used in collaborative mean estimation. 

\medskip

\noindent {Confidence intervals for special cases are as follows:}
\begin{enumerate} \item \textit{Gaussian distribution:}  If the local mean $\bar{x}_{a,a}(t)$ can be considered as a Gaussian distributed random variable,  then it is distributed as 
	$\bar{x}_{a,a}(t) \sim \mathcal{N}(\mu_a, \sigma \frac{1}{\sqrt{t}}).$ For this distribution we can easily calculate the confidence intervals for a given $\delta$ as 
	\begin{equation}\sigma B_{\delta}(t) = \beta_{\delta}(t)=z_{1-\delta /2} \frac{\sigma} {\sqrt(t)},\end{equation} where the value $z_{1-\delta /2}$ for several commonly used values $(1-\delta, z_{1-\delta /2})$ for various $1-\delta$ are given next: $(0.95, 1,96), (0.975, 2.24), (0.99, 2.58), (0.999, 3.29).$
	
	\item \textit{Student's t-distribution:} If the number of samples is small and the standard deviation is also estimated by $\hat{\sigma}$, then we should use the Student's t-distribution, with the confidence bounds defined by:
	$
		\beta_{\delta}(t) = \sigma B_{\delta}(t)  = t_{1 - \delta/2, \nu} \frac{\hat{\sigma}}{\sqrt{t}},
	$
	where the value $t_{1-\delta/2, \nu}$ depends on $\delta$ and the degree of freedom $\nu$. 
	\item \textit{$\sigma$-sub-Gaussian distribution:} The $\sigma$-sub-Gaussian distribution is a class of probability distributions characterized by tails that are no heavier than those of a Gaussian distribution with variance $\sigma$. If $x_{a}(t)$ is a random variable distributed as $\sigma$-sub-Gaussian, then:
	\begin{equation}\mathbb{E}[e^{\lambda(\mathrm{x}-\mu_a)}] \le e^{\frac{\sigma^2 \lambda^2}{2}}, \quad \text{ for all } \lambda \in \mathbb{R},\end{equation}
	where the random variable is indicated by $\mathrm{x}$. For this distribution holds:
	\begin{equation}\mathbb{P}(|\mathrm{x} - \mu_a| \ge t) \le 2e^{-\frac{t^2}{2\sigma^2}}, \quad  \text{ for all } t \ge 0.\end{equation}
	These bounds show the rapid decay of the probability of large deviations from the mean. For variance $\sigma$-sub-Gaussian distributed random variable we have $\mathrm{Var}(\mathrm{x} ) \le \sigma^2$.
	
	Examples of $\sigma$-sub-Gaussian distributions are: 
	\begin{itemize}
		\item Gaussian distribution with $\mathrm{x} \sim \mathcal{N}(\mu_a, \sigma_1)$, for $\sigma_1 \le \sigma$;
		
		\item Rademacher distribution with a random variable $\mathrm{x}$ taking values +1 and -1 with equal probability is 1-sub-Gaussian; 
		
		\item Uniform distribution within $[-a, a]$ is $\frac{a}{\sqrt{3}}$-sub-Gaussian distributions; 
		
		\item Symmetric Bernoulli distribution with a random variable taking values $+a$ and $-a$ with equal probability is $a$-sub-Gaussian.
		
		\item Linear combinations of independent $\sigma_i$ sub-Gaussian distributions are $\sqrt{a_i\sigma^2_i}$ sub-Gaussian distribution. 
		
		\item Vectors and matrices with independent $\sigma$-sub-Gaussian distributed elements are $\sigma$-sub-Gaussian distributed. 
		
	\end{itemize}

The confidence intervals (by Laplace method, a consequence of Maillard  Lemma) are defined using \cite{maillard2019mathematics, asadi2022collaborative,galante2025scalable}
\begin{equation}\mathbb{P}\Big(\bar{x}_{a,a}(t)-\mu_a\Big) >\sigma \sqrt{\frac{2}{t}\Big(1+\frac{1}{t}\Big)\ln\Big(\frac{\sqrt {t+1}}{\delta/2}\Big)} \Big)\end{equation}
as the Laplace bound
\begin{equation}\beta_{\delta}(t)=\sigma B_{\delta}(t) =\sigma \sqrt{\frac{2}{t}\Big(1+\frac{1}{t}\Big)\ln\Big(\frac{\sqrt {t+1}}{\delta/2}\Big)}.\end{equation}
\item \textit{Bounded Fourth-Moment Distributions:} Consider a distribution with bounded fourth-moment,
\begin{equation}\mathbb{E}\{ (x_a(t)-\mu_a)^4\}\le \mu_4.\end{equation}
The true mean, $\mu_a$, with probability greater than $1-\delta$, belongs to the confidence intervals centered in $\bar{x}_{a,a}(t)$ and bounded with  \cite{galante2025scalable} 
\begin{equation}\beta_{\delta}(t)=\sigma \sqrt[4]{\frac{2(\kappa +3)}{\delta/2}\frac{(1+\ln^2(t))}{t}}. \end{equation}
When all the variables are identically distributed, $\kappa$ corresponds to the kurtosis 
\begin{equation}\kappa=\frac{\mathbb{E}\{(\mathrm{x}-\mu_a)^4\}}{\sigma^4}.\end{equation}
This result produces tighter confidence intervals than the one obtained using Laplace bounds. For Gaussian distribute $\bar{x}_{a,a}(t)$ we have $\kappa=3$. For Gaussian distribution and $\delta=0.01$ we can write  $\beta_{\delta}(t)=8.32 \sigma \sqrt[4]{(1+\ln ^2(t))/t}$.

\end{enumerate}

Since confidence intervals will play a crucial role in the convergence of the presented approaches, we will discuss their properties and behavior in more detail. 

\bigskip

\noindent\textbf{The confidence interval properties:} 
The confidence intervals contain the expected true value $\mu_a$ of the data $\bar{x}_{a,a}(t)$ with probability higher than $1-\delta$. They also tend to 0 as $t \to \infty$. Therefore, for sufficiently large $t$:
\begin{itemize}
\item 
Two confidence intervals of agents $a$ and $a'$ of the same similarity class will intersect with high probability, since they contain at least one common point, the true expected value, $\mu_a$. 
\item 
Two confidence intervals of agents $a$ and $a'$ of different similarity classes will not intersect with a high probability, since they are centered around different true expected values, $\mu_a \ne \mu_{a'}$ and the widths of the confidence intervals tends to 0. 
\end{itemize}

So, the \textit{intersection of confidence intervals} will work as an indicator if the agents considered $a'$ should be considered to belong to the similarity class of $a$ or not. This criterion is used for similarity class determination. 

Initially, all neighbors are assumed to belong to the similarity class of $a$. After each instant $t$, all agents first check the confidence intervals with each neighbor. 
\begin{itemize}
\item
If the confidence intervals intersect, that neighbor is kept in its neighborhood. 
\item 
If the confidence intervals do not intersect, then the edge connecting $a$ and $a'$ is disconnected and that agent is no longer considered as a neighbor (value of 1 in the adjacency matrix for that connection is set to 0 and we have a new adjacency matrix).  This is why the adjacency matrix, defining the agent neighborhoods, is time dependent $\mathbf{A}(t)$.
\end{itemize}

\subsection{Expected Separation Time}\label{Sec:ExSep}

Consider two agents $a$ and $b$ from two different similarity classes with mean values $\mu_a$ and $\mu_b$. Without loss of generality assume that $\mu_a>\mu_b$, that is $\Delta=\mu_a-\mu_b>0$.
Their local means at time $t$ are denoted by $\bar{x}_{aa}(t)$ and $\bar{x}_{bb}(t)$. The critical situation for the confidence intervals is when 
\begin{equation}\beta_{\delta}(t)+\beta_{\delta}(t-1)=\bar{x}_{aa}(t)-\bar{x}_{bb}(t).\end{equation}
Assuming that $\beta_{\delta}(t)\approx\beta_{\delta}(t-1)$ and taking the expected values of the previous equation, we get:
\begin{equation}2\beta_{\delta}(t)=\mu_a-\mu_b.\end{equation}
By solving this equation for $t$, with a given $\mu_a-\mu_b$, we get the expected separation time $T_{sep}$. In general, if there are more than two classes, then the expected separation time is obtained by solving $2\beta_{\delta}(t)=\min|\mu_a-\mu_b|=\min\{\Delta_{ab}\}$ for all $\mu_a$ and $\mu_b$.

\section{Sample Variance Estimation}\label{Sec:VarEst}

In order to calculate the confidence intervals, we have to know the standard deviation $\sigma$ of the data $x_a(t)$. In literature on colME, \cite{asadi2022collaborative, galante2025scalable}, it has been assumed that the variance(s) (standard deviation) of random distributions are known. In reality that is not the case and the variance has to be estimated. In addition the estimation has to be online, and may or may not include collaboration. The variance estimation is mean dependent, and since the mean is the final goal of our problem, it would be desirable to have a mean-invariant estimator of the variances.  The online, local and collaborative, mean-invariant estimation of the variance is the first goal of this paper, that will be presented next.  

To estimate the variance, we will define a new random variable
\begin{equation}d_a(t)=x_a(t)-x_a(t-1).\end{equation}
Note that the new random variable $d_a(t)$ has zero mean for all $a \in \mathcal{A}$, $\mathbb{E}\{d_a(t)\}=0$, since we assumed that the exact expected value does not change over time for a given agent, $\mathbb{E}\{x_a(t)\}=\mu_a$. 
\textit{It means that for $d_a(t)$ all agents belong to the same similarity class, with a known zero mean. }

We can efficiently estimate the variance from just several initial data samples (in the case of the same variance, all agents can collaborate in the variance estimation, since the estimation is mean-invariant). 

The variance estimation is based on the fact that
\begin{gather}\mathbb{E}\{d^2_a(t)\}=\mathbb{E}\{\Big(x_a(t)-x_a(t-1)\Big)^2\}\\
	=\mathbb{E}\{\Big(x_a(t)-\mu_a-(x_a(t-1)-\mu_a)\Big)^2\} \\
	= \mathbb{E}\{(x_a(t)-\mu_a)^2\} +\mathbb{E}\{(x_a(t-1)-\mu_a)^2\}=2 \sigma ^2,
\end{gather}
\noindent since $\mathbb{E}\{x_a(t)-\mu_a\}\mathbb{E}\{x_a(t-1)-\mu_a\}=0$.

\bigskip

In practical implementation, we will consider two methods:

\noindent (i) One is based on using local data, without collaboration with neighborhood. This method will be used later in combination with local means. Then the estimated variance (and standard deviation as its squared root) for an agent $a$, at an instant $t$, is 
\begin{equation}\hat{\sigma}_a^2(t)=\frac{1}{2t}\sum_{\tau=1}^t \big(x_a(\tau)-x_a(\tau-1)\big)^2.\end{equation}
Since this method does not use collaboration, it can (and will) be used for estimation when each similarity class has different variance and for defining the confidence intervals for sample standard deviation.    

\noindent (ii)  The other, which assumes common inter-agent communication within the defined neighborhood only.  It is of practical importance, as it does not require any additional communication links.

Agents can communicate only within their defined neighborhood (denoted by $\mathcal{N}_a$) over the established, initially $r$ links. The agents exchange (depending on the collaboration method used) the local means or the local sums. In both cases, we can recover the most recent data values $x_a(t)$. Assuming that the local means are exchanged, at $t=0$, the data values are obtained as
\begin{equation}m_{a,a}(0)=x_a(0), \text{ and } m_{a,a'}(0)=x_{a'}(0), \text{ for } a' \in  \mathcal{N}_a.\end{equation} 
At $t=1$ we have 
\begin{equation}m_{a,a}(1)=x_a(0)+x_a(1), \  m_{a,a'}(1)=x_{a'}(0)+x_{a'}(1), \  a' \in  \mathcal{N}_a.\end{equation}
Then we can form $r$ differences (for all $r$ neighboring nodes) using $m_{a,a'}(1)-2m_{a,a'}(0)$ as
\begin{equation}x_a(1)-x_a(0), \text{ and } x_{a'}(1)-x_{a'}(0), \text{ for } a' \in  \mathcal{N}_a.\end{equation}
We can estimate the variance using these $r$ values. However, the number of differences may not be sufficient for an accurate estimate, as $r$ is usually small. Then, this procedure should be repeated for several initial instants $t$ in order to accumulate enough values for a satisfactory estimation.
For example, for each $t=2,3\dots,10=T_s$ we form $r$ differences:
\begin{gather*}x_a(2)-x_a(1), \text{ and } x_{a'}(2)-x_{a'}(1), \text{ for } a' \in  \mathcal{N}_a, \\
	x_a(3)-x_a(2), \text{ and } x_{a'}(3)-x_{a'}(2), \text{ for } a' \in  \mathcal{N}_a, 
	\\
	\dots \\
	x_a(10)-x_a(9), \text{ and } x_{a'}(10)-x_{a'}(9), \text{ for } a' \in  \mathcal{N}_a,
\end{gather*}
and estimate 
\begin{equation}\hat{\sigma}^2_a (T_s)= \frac{1}{2T_s(|\mathcal{N}_a(T_s)|+1)}\sum_{\substack{a' \in \mathcal{N}_a(0) \cup \{a\} \\ t=1,\dots,T_s}} \big(x_{a'}(t) - x_{a'}(t-1)\big)^2.
\end{equation}

Note again that in the variance estimation process, all $A$ agents in the system can collaborate in the variance estimation if \textit{the variance of all agents is the same and classes (the data distributions)} $D_a$ \textit{ differ only in the mean value}. So, all these values are squared and their mean is found. If we use, for example, 10 instants and $12$ neighbors $a'$, then we get 120 + 12 differences (including the central node $a$).

\textit{During the initial instants, until we estimate variance,}  we keep it very large so that all confidence intervals intersect (as they usually do initially), $\sigma_a=10$.

\noindent \textbf{Local Sums Case}: If the agents exchange sums,then 
\begin{equation}x_{a'}(t)=m_{a,a'}(t)-m_{a,a'}(t-1) =\sum_{\tau=1}^{t}x_{a'}(\tau)-\sum_{\tau=1}^{t-1}x_{a'}(\tau).\end{equation}

\noindent \textbf{Local Means Case}: Agents may exchange collaborative local means $\bar{x}_{aa}(t)$.  Then,  for $t \le T_s$, we calculate the data values as    
\begin{equation}x_{a'}(t)=t\bar{x}_{a,a'}(t)-(t-1)\bar{x}_{a,a'}(t-1)\end{equation}
with $\bar{x}_{a,a'}(t)=\mu_a(t)$ for $t \le T_s$.

If the mean estimation algorithm was defined in such a way that the collaborative means were only exchanged, then to avoid the influence of neighboring agents on the exchanged means, during variance estimation, and to ensure that only local means are exchanged, in consensus-based approaches we can use $\alpha(t)=0$ for $t \le T_s$ and the given value $\alpha(t)$ afterwords. This will also avoid agent collaboration in mean calculation during the initial phase, when the neighbors selection is highly unreliable. 

\bigskip 

This procedure is summarized within the Algorithm \ref{AlgSE}.

\begin{algorithm}[H]
	\caption{Local Collaborative Estimation of the Standard Deviation,  $\sigma$, using Agent Neighborhood}
	\label{AlgSE}
	\begin{algorithmic}[1]
		\For{each \( t = 0 \) to \( T \)}
		\For{each agent \( a \in \mathcal{A} \)} 
		\State \( s \gets 0, \ \ \ \hat{\sigma}_a \gets 10 \) (a large value so that CI intersect)
		\If{\( t \le T_s\)} (such that $T_s|\mathcal{N}_a| \sim 100$)
		\State Update \( s \):
		\[
		s \gets s + \sum_{a' \in \mathcal{N}_a \cup \{a\}} |x_{a'}(t) - x_{a'}(t-1)|^2
		\]
		\ElsIf{\( t = T_s \)}
		\State Compute \( \hat{\sigma}_a \):
		\[
		\hat{\sigma}_a \gets \sqrt{\frac{s}{2T_s ( |\mathcal{N}_a(0)|+1)}}
		\]
		\EndIf
		\EndFor
		\State Compute Confidence Intervals and Collaborative Mean
		\EndFor
	\end{algorithmic}
\end{algorithm}

\subsection{Central Fourth Moment and Kurtosis Estimation}

For confidence intervals based on the fourth-order moment we have to estimate the central fourth-order moment
\begin{equation}\mu_4 = \mathbb{E}\{(\mathrm{x}-\mu_a)^4\}\end{equation}
Again, it will be done using the property that 
\begin{gather*}\mathbb{E}\{(x_a(t)-(x_a(t-1))^4\} = \mathbb{E}\{(x_a(t)-\mu_a-(x_a(t-1)-\mu_a))^4\}  \\ = \mathbb{E}\{(x_a(t)-\mu_a)^4\}+6\mathbb{E}\{(x_a(t)-\mu_a)^2\}\mathbb{E}\{(x_a(t-1)-\mu_a)^4\} \\ +\mathbb{E}\{(x_a(t-1)-\mu_a)^4\} 
	= \mu_4+ 6\sigma^2 \sigma^2 +  \mu_4 
\end{gather*}
for symmetric distribution of data with the mean value not changing over time.
Therefore, the estimation can be done using:
\begin{gather*}
	\mu_4= \frac{1}{2}\mathbb{E}\{(x_a(t)-(x_a(t-1))^4\}- 3 \sigma^4 \\
	\hat{\mu}_{4a} = \frac{1}{2T_s(|\mathcal{N}(t)|+1}\sum_{\substack{a' \in \mathcal{N}_a(0) \cup \{a\} \\ \tau=1,\dots,t}} \big(x_{a'}(\tau) - x_{a'}(\tau-1)\big)^4 -3 \hat{\sigma}^4_a \\
	\hat{\kappa}_a = \frac{\hat{\mu}_{4a}}{\hat{\sigma}^4_a} \\ =\frac{1}{2T_s(|\mathcal{N}(t)|+1)\hat{\sigma}^4_a}\sum_{\substack{a' \in \mathcal{N}_a(0) \cup \{a\} \\ \tau=1,\dots,t}} \big(x_{a'}(\tau) - x_{a'}(\tau-1)\big)^4 -3.
\end{gather*}

For any Gaussian distribution, kurtosis is constant, $\kappa=3$. The kurtosis in statistics is also used as a measure of the deviation of the distribution from the Gaussian distribution, using excess kurtosis $\kappa - 3$, which is always zero if the distribution is Gaussian.  Distributions with high kurtosis have heavy tails. Kurtosis of a distribution smaller than 3 means lighter tails compared to a normal distribution.

\section{Review of Collaborative Mean Estimation Approaches with Estimated Variance}
The basic approach (colMe) uses a communication within the whole set of agents for collaborative processing \cite{adi2020machine}. However, this approach faces computational and memory complexity of order $A$, which is not practical for large systems.  In order to reduce the system complexity, a graph can be assigned to the given set of agents as nodes. Then we start with a random graph with a defined starting degree $r$ of each node (that is, establishing r links for communication with neighbors for each agent) as in B-colME and C-colME. All these approaches will shortly be reviewed next.
\subsection{ColME}
The problem of estimating the mean within a scenario involving multiple agents is considered in~\cite{asadi2022collaborative}. Each agent in the group $(\mathcal{A} = \{1, 2, \ldots, A\})$ is tasked with determining the mean $\mu_a$ of its distribution. It has been assumed that each distribution is $\sigma$-sub-Gaussian. The framework is both online and collaborative: 
\begin{itemize}
	\item Each agent receives data in sequence from its distribution.	
	\item Each agent can interact with one agent to exchange data. 
\end{itemize}

These agents function similarly to various user devices in a network that operate simultaneously, allowing each to collect a new sample and communicate with another agent at each time instance.

In a more formal sense, we assume synchronized timings between agents, where at each discrete time, $t=0,1,2,\dots,T$, an agent $a$:
\begin{itemize}
	\item
	Obtains a new sample $x_a(t)$ from its distribution with mean $\mu_a$, and updates its local mean estimate, given by $\bar{x}_{aa}(t)$. 
	\item The agent then selects another agent $a'$ for querying. In return for its query, agent $a$ receives the local mean $\bar{x}_{a'a'}^t$ from agent $a'$, representing the aggregated information of $a'$ of its independent samples, and records it in its own dataset as $\bar{x}_{a,a'}(t)$, in conjunction with the sample count $n_{a,a' }(t) = t$. The agent $a'$ can also send the aggregated sum of its local samples $m_{a'a'}^t$ stored as $m_{aa'}^t$, together with $n_{a,a' }(t) = t$, with $\bar{x}_{a,a'}(t)=m_{a,a'}(t)/n_{a,a' }(t)$ or $m_{a,a'}(t)=n_{a,a' }(t)\bar{x}_{a,a'}(t)$.
	\item Each agent keeps a log of all recent local averages, \begin{equation}[\bar{x}_{a,1}(t), n_{a,1}(t), \ldots, \bar{x}_{a,A}(t), n_{a,A}(t)]\end{equation} obtained from other agents, including its own local average. This information is used to update the mean estimate of each agent at time $t$,
	\begin{equation}\mu_a(t)=\frac{\sum_{a'}m_{a,a'}(t)}{\sum_{a'}n_{a,a'}(t)}=\frac{n_{a,a'}(t)\sum_{a'}\bar{x}_{a,a'}(t)}{\sum_{a'}n_{a,a'}(t)}\end{equation}
	
\end{itemize}

When a query occurs, instead of acquiring just one sample, as in multi-armed bandit problems, the querying agent gains comprehensive insight into accumulated observations up to the current time. This method yields more extensive data than conventional bandit settings.

In practice, each agent $a$ assumes that its initial neighborhood $\mathcal{N}a(0)$ is the whole set of other agents
	\begin{gather*}
		\mathcal{N}_0(0) = [1,2,3,\dots,A],   \\
		\mathcal{N}_1(0) = [2,3,4,\dots,A, 0],  \\
		\dots,  \\ 
		\mathcal{N}_A(0) = [0,1,2,\dots,A-1].
	\end{gather*}
At instant $t=0$ agent $a$ visits agent $a+1$ (the first element in the list), exchanges data, and decides if the visited agent is within the same similarity class or not (checking the confidence interval). If it is, then the neighborhood list is just roll-rotated left keeping all agents in the list, and the list
\begin{equation}[\bar{x}_{a,1}(t), n_{a,1}(t), \ldots, (\bar{x}_{a,A}(t), n_{a,A}(t)])\end{equation}
is updated with a new mean and number of terms.
If it is not, then the first agent in the neighborhood list is deleted, and the remaining list is kept as it is (\textit{Restricted-Round-Robin} approach) with the corresponding data being deleted (or set to zero) in the data list.

\begin{algorithm}
	\caption{Mean Estimation using colME}
	\begin{algorithmic}[1]
		\State \textbf{Input:}   $\delta$, $\epsilon $, Distribution $D_a$, 
		\State Initialize ordered neighborhood $\mathcal{N}_a$ for all nodes $a$
		\State Initialize exchanged local means and times list $(\bar{x}_{aa'}(0), n_{aa'}(0))$ for all $a$ and $a'$
		\For{$t = 1$ to $T$}
		\ForAll{nodes $a \in \mathcal{A}$ \textbf{in parallel}}
		\State Draw $x_a(t)$ from the data distribution
		\State Calculate local mean $\bar{x}_{aa}(t)$
		\State Calculate confidence intervals for $a$ and the first agent $ a' =\mathcal{N}_a[0]$ using $\bar{x}_{aa}(t)$ and $\bar{x}_{a'a'}(t-1)$
		\If{confidence intervals for $a$ and $a'$ do not intersect}
		\State $\mathcal{N}_a[0] = [ \ ]$, $\bar{x}_{aa'}(t)=0$, and $n_{aa'}(t)=0$
		\Else
		\State Roll-rotate left  $\mathcal{N}_a$
		\State Updated list of local means and times in $a$ by $\bar{x}_{aa'}(t)$ and $n_{aa'}(t)=t$.  
		\EndIf
		\State \textbf{Output:} Collaborated mean for agent $a$, calculated as a sum over all $a'$ of $n_{aa'}(t)\bar{x}_{aa'}(t)$ divided by the sum of $n_{aa'}(t)$.
		\EndFor
		\EndFor
	\end{algorithmic}
\end{algorithm}

\subsection{Graph-Based  C-colME and B-colME}

In large-scale systems, having agents query all other agents or maintain a memory that scales linearly with the number of agents $A$ can be infeasible. Practically, each agent can instead focus on a limited subset of agents that is manageable in size, which will be done using a graph as the domain next.

\begin{figure}[htp]
	\centering
	\includegraphics[scale=.35]{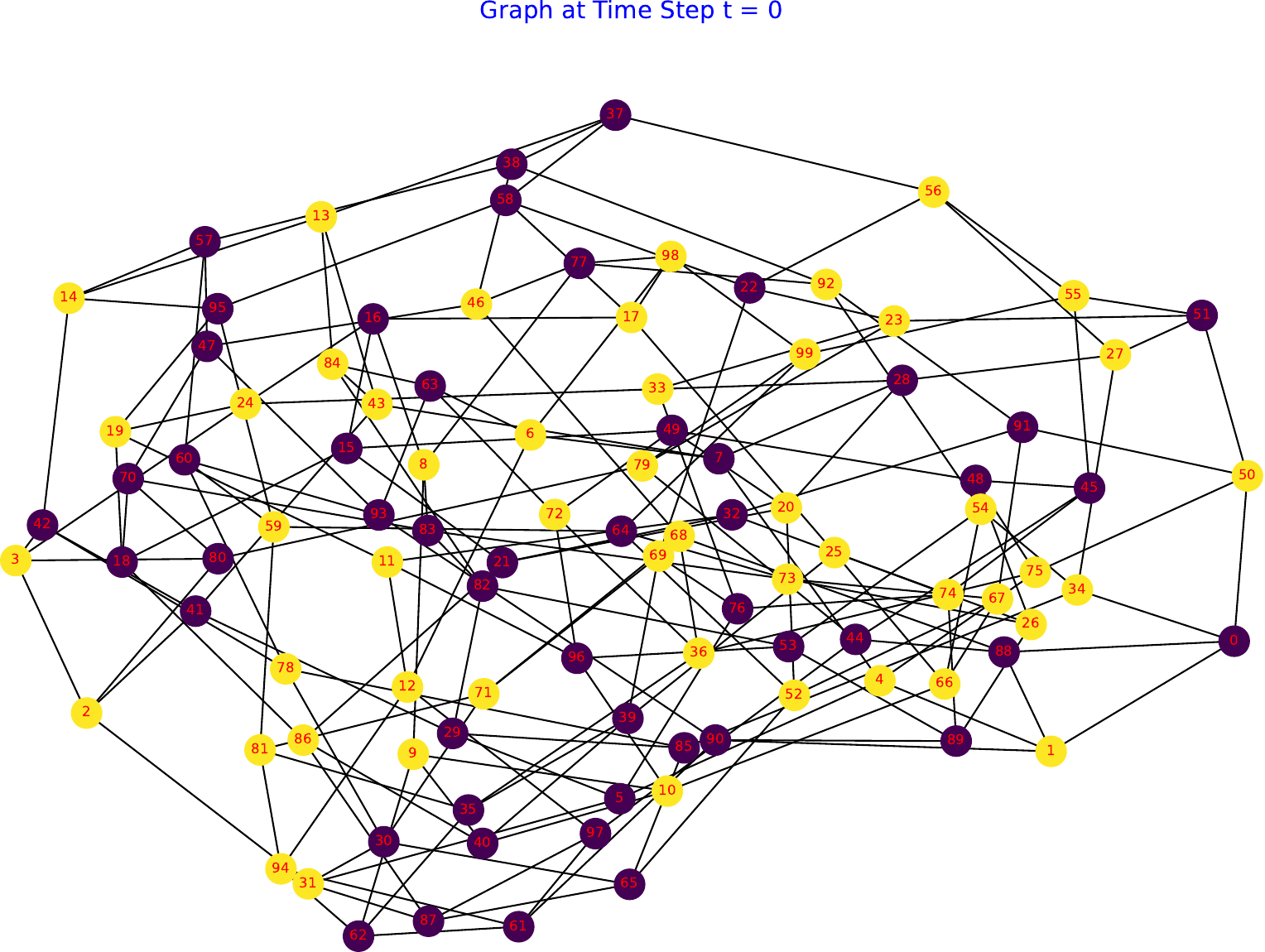}
	\caption{Random regular graph with $A=100$ agents/nodes and $C=2$ classes, arbitrary connected, at $t=0$, with $r=4$ links using a random regular graph $\mathcal{G}(100, 4)$ .}
	\label{Fig_Ill}
\end{figure}

The random graph (agents with assumed links) is denoted as $\mathcal{G}(\mathcal{A}, \mathcal{E})$, where $\mathcal{A}$ is the set of nodes/agents and $\mathcal{E}$ are the edges, indicating the communication links between the nodes.  In particular, the initial graph will be a random sample of a random graph $\mathcal{G}(A, r)$ with exactly $r$ edges from each node at $t=0$. The selection of $r$ is very important and has bee discussed in detail in \cite{galante2025scalable}, Fig. \ref{Fig_Ill}.

Since we do not now know anything about the class, many of the connections are wrong connections, meaning that agents from different classes established communication links. 
The task is as follows.
\begin{itemize}
	\item To eliminate wrong links over time and to leave only the links among the same class agents.
	\item Define a good formula that will use the means and other data from neighbors to improve the estimation.
\end{itemize}

\bigskip 

\noindent{\bf The consensus-Based Approach (C-colME)} uses both the local means $\bar{x}_{a,a}(t)= \frac{1}{t}\sum_{\tau=1}^{t}x_a(\tau)$, or in vector form, for all agents $\mathbf{X}(t)$, and collaborative means calculated using the estimated similarity class, within the random neighborhood $\boldsymbol{\mu}(t)$. Note that if the local sums are calculated in the code, then the local mean vector is $ \mathbf{X}(t)= \mathbf{M}(t)/t$.

We start with $\boldsymbol{\mu}(0)= \mathbf{X}(0)$ and calculate all next consensus values combining local means $\mathbf{X}(t)$, with weight $(1-\alpha(t))$ and previous consensus mean $\boldsymbol{\mu}(t-1)$ with weight $\alpha(t)$, that is
\begin{equation}\boldsymbol{\mu}(t+1)=(1-\alpha(t))\mathbf{X}(t)+\alpha(t)\mathbf{W}(t)\boldsymbol{\mu}(t)\end{equation}
where $\mathbf{W}(t)$ is the weight matrix that will be discussed next.

\begin{itemize}
	\item
	In order to obtain an unbiased result, we must obviously have the sum of the rows of the matrix $\mathbf{W}_t$ to be one. If we assume that all means are equal to 1, then $\boldsymbol{1}=(1-\alpha(t))\boldsymbol{1}+\alpha(t)\mathbf{W}(t)\boldsymbol{1}$, should hold, which means that all rows of $\mathbf{W}$ should sum to 1. Since this matrix is symmetric, this holds for columns as well, and such matrix is called a double stochastic matrix. 
	
	\item At the same time, the matrix $\mathbf{W}(t)$ should play the role of the adjacency matrix $\mathbf{A}(t)$ (weighted adjacency matrix) since it should combine the values of $\boldsymbol{\mu}(t-1)$ within the neighborhood of a given node only, hoping that this neighborhood would be a part of the similarity class (after some time steps).
	
	\item We will assume that the values of matrix $\mathbf{W}(t)$ are constant for each node, and take their value as 
	\begin{equation}W_{i,j}(t)=W_{j,i}(t)=\frac{1}{\max \{D_i,D_j\}+1} \text{ for } A_{i,j}(t)=1\end{equation}
	and $W_{i,j}(t)=0$ for $A_{i,j}=0$. The values $D(i)$ are degrees of matrix $\mathbf{A}(t)$, that is $D(i)=\sum_jA_{i,j}(t)$
	
	\item In order to achiever that $\sum_jW_{i,j}(t)=1$ we have 
	\begin{equation}W_{i,i}(t)=1-\sum_jW_{i,j}(t)\end{equation}
	
\end{itemize}

\begin{algorithm}
	\caption{Mean Estimation using C-colME with Variance Estimation}
	\begin{algorithmic}[1]
		\State \textbf{Input:}  Graph $G(\mathcal{A}, \mathcal{E})$, with matrix $\mathbf{A}(0)$, $\delta$, $\epsilon $, Distribution $D_a$, 
		\State Initialize neighborhood $\mathcal{N}_a(0)$ for all nodes $a \in \mathcal{A}$, $s=0$, initial $\hat{\sigma}(a)=10$
		\For{$t = 1$ to $T$}
		\ForAll{nodes $a \in \mathcal{A}$ \textbf{in parallel}}
		\State Draw $x_a(t)$ from the data distribution
		\State Calculate local mean $\bar{x}_{aa}(t)$
		\State Calculate confidence intervals for all $a' \in \mathcal{N}_a(t)$ using $\bar{x}_{aa}(t)$ and $\bar{x}_{a'a'}(t-1)$
		\If {$t \le T_s$ } 
		\State Update $s \leftarrow s + (x_{a'}(t)-x_{a'}(t-1))^2$ for $a' \in \mathcal{N}_a(t)\cup\{a\}$
		\ElsIf{ $t == T_s$ } 
		\State Standard deviation $\hat{\sigma}(a)= \sqrt{\frac{s}{T_s(\mathcal{N}_a+1)}} $ 
		\EndIf    
		\If{confidence intervals for $a$ and $a'$ do not intersect}
		\State $\mathcal{N}_a(t) = \mathcal{N}_a(t)\setminus\{a'\}$,  that is,
		\State Update the graph adjacency matrix, $A_{aa'}(t)= A_{a'a}(t)=0$
		\State Calculate double stochastic matrix, $\mathbf{W}(t)$.  
		\EndIf
		\State Exchange local sums/collaborative means/messages with neighbors
		\State \textbf{Output:} Collaborated mean for agent $a$ according to \State $\boldsymbol{\mu}(t+1)=(1-\alpha(t))\mathbf{X}(t)+\alpha(t)\mathbf{W}(t)\boldsymbol{\mu}(t)$ 
		\EndFor
		\EndFor
	\end{algorithmic}
\end{algorithm}

It is easy to show that the C-colME converges to oracle solution, \cite{masterwithcode}. After pruning is completed  by the confidence interval checks at some $t=t_0$ and $\alpha(t)$ reaches 1, we have:
\[
\boldsymbol{\mu}(t+1)={\color{gray!80}(1-\alpha(t))\mathbf{X}(t)+\alpha(t)}\mathbf{W}(t)\boldsymbol{\mu}(t) \to \mathbf{W}^ {t-t_0}(t_0)\boldsymbol{\mu}(t_0).
\]
For  double stochastic matrix, its power  $\mathbf{W}^ k$ tends to a fully connected adjacency matrix on each graph component, as $t$ increases. 

\bigskip 

\noindent{\bf The Message-Passing Algorithm (B-colME)} approach, the agents form an exchange message of dimension $d \times 2$, where $d$ is the distance in the graph (a parameter). Each agent first calculates a sum of its own samples up to the actual time instant $t$
\begin{equation}
	m_{1,1}^a(t) = \sum_{\tau=1}^{t}x_a(\tau).
\end{equation}
These sums for all agents $a \in \mathcal{A}$  can be considered as elements of a column vector denoted by $\mathbf{M}(t)$. The sums are exchanged as part of the message (in particular, this is \textit{the first row in the message table}, along with the number of samples in the sum $t$, as the second element of the first row in the message). Before continuing with the next rows, we will state that these messages are exchanged among neighboring vertices. A vertex $a'$ from the neighborhood of $a$, denoted by $\mathcal{N}_a$, sends this information 
\begin{equation}
	m_{1,1}^{a'\to a}(t) = \sum_{\tau=1}^{t}x_{a'}(\tau).
\end{equation}
Notice that the second element of the first row, $	m_{1,2}^{a'\to a}(t)=t$, is the number of samples.  

The node $a$ accumulates all values from the messages from $\mathcal{N}_a$ as $\sum_{a' \in \mathcal{N}_a} m_{1,1}^{a'\to a}(t)$. 

In matrix form sum of all sums over neighbors, for agents,  can be written for the whole system of agents as 
\begin{equation}\mathbf{M_1}(t)=\mathbf{A}(t)\mathbf{M}(t)\end{equation}
since the symmetric adjacency matrix of the graph $\mathcal{G}(\mathcal{A}, \mathcal{E})$, denoted at an instant $t$ by $\mathbf{A}(t)$ contains 1s in the row corresponding to the adjacent agents to each node and by multiplying the matrix $\mathbf{A}(t)$ with vector $\mathbf{M}(t)$ we exactly sum the sums of all neighboring agents to each node, that is, we sum the first elements, $m_{1,1}^{a'\to a}(t)$ in the first row of the exchange matrix for all neighbors of each $a$. 

The number of elements in the sum $\mathbf{M_1}(t)$ for each agent $a$ is equal to the number of its neighbors multiplied by $t$. The number of neighbors for each agent is equal to the sum of matrix $\mathbf{A}(t)$ over its rows or columns (symmetric matrix), that is $\mathbf{D1}(t)=[\sum_j A_{i,j}(t), i=1,2,,\dots,n]$

Next we consider the \textit{second row of the message table}. This row contains the sums of the first neighbors of neighbors of an agent $a$. The first element in the second row of the exchange messages is 
\begin{equation}
	m_{2,1}^{a'\to a}(t) = \sum_{a'' \in \mathcal{N}_{a'}, a ''\ne a}m_{1, 1}^{a'' \to a'}(t-1).
\end{equation}
This element of the exchange message gives the sum of the sums of agents at distance 2 in the graph of the agent considered, $a$. These sums will be summed for all $a' \in \mathcal{N}_{a}$ and used along with the total number of terms (contained in the second element of the second row of the message table) to estimate the collaborative mean.

\section{Two-Fold Confidence Intervals for Standard Deviation and Mean}

Instead of presenting results obtained with colME, B-colME, and C-colME with estimated variance (that may be found in \cite{masterwithcode}) we will go to an extended, more complex case with two-fold and three-fold confidence intervals (not considered in the literature) using the estimated sample variance. 

The confidence intervals based solely on the local means are considered in existing works on colME. In the cases where \textbf{the true means are very close and the standard deviations differ more significantly}, the convergence of separation based on the mean-based confidence intervals would be extremely slow. However, we can add into consideration the standard deviation estimates and their confidence intervals, forming two-fold confidence intervals. 

The estimated standard deviation at an instant $t$, for each agent, is already defined in Section \ref{Sec:VarEst}. Its local value, using only the considered agent $a$, is   
\begin{gather}\hat{\sigma}_a(t) = \sqrt{\frac{1}{2t}\sum_{\tau=1}^t\big(x_{a}(\tau) - x_{a}(\tau-1)\big)^2}\\
	\hat{\sigma}^2_a(t) = \frac{1}{2t}\sum_{\tau=1}^t\big(x_{a}(\tau) - x_{a}(\tau-1)\big)^2= \frac{1}{2t}\sum_{\tau=1}^t d_{a}^2(\tau).
\end{gather}
The statistics of the standard deviation for the sub-Gaussian data is quite complex. 
To estimate the variance of the sample variance estimator, recall that for a zero-mean $\sigma$-sub Gaussian random variable $\mathbf{x}$, holds $\mathbb{E}\{\mathbf{x}^4\} \leq C\sigma^4$ for some constant $C$ that depends on the specific distribution. For a Gaussian distribution, $C = 3$. For the distributions when the kurtosis exists we can write $\mathbb{E}\{\mathbf{x}^4\} = \kappa_x \sigma^4$.

First, consider the variance of $d^2_a(t)=(x_a(t)-x_a(t-1))^2$. Using $\mathrm{Var}(\mathbf{x})=\mathbb{E}(\mathbf{x}^2)-(\mathbb{E}(\mathbf{x}))^2$,  for $d_a^2(t)$ we can write
\begin{gather}\mathrm{Var}(d_a^2(t)) = \mathbb{E}\{d_a^4(t)\} - (\mathbb{E}\{d_a^2(t)\})^2 =\kappa_d \sigma_d^4 - \sigma_d^4 \\ 
	= (\kappa_d-1)\sigma_d^4=(\kappa_d-1) (2\sigma^2)^2.\end{gather}
Now, we estimate the variance of $\hat{\sigma}^2_a(t)$ as
\begin{gather}\mathrm{Var}(\hat{\sigma}^2_a(t)) = \mathrm{Var}(\frac{1}{2t} \sum_{\tau=1}^{t} (x_a(\tau)-x_a(\tau-1))^2) \\= \frac{1}{4t^2} \mathrm{Var}(\sum_{\tau=1}^{t} (x_a(\tau)-x_a(\tau-1))^2) \\
	= \frac{1}{4t^2} \sum_{\tau=1}^{t} (\kappa_d-1) (2\sigma^2)^2=  \frac{(\kappa_d-1)\sigma^4}{t}.
\end{gather}
To ensure statistical independence of two values of $d_a(t)=x_a(t)-x_a(t-1)$ we can use $t=1,3, \dots$. However, using every $t$ produced the same statistical results. 

Estimating the variance of the sample standard deviation involves the Delta method,  for $g(x) = \sqrt{x}$, with $g'(x) = \frac{1}{2\sqrt{x}}$. Using this method we get
\begin{gather}\mathrm{Var}(g(\hat{\sigma}^2)) \! \approx \! [g'(\sigma^2)]^2 \mathrm{Var}(\hat{\sigma}^2) \! = \! (\frac{1}{2\sqrt{\sigma^2}})^2 \mathrm{Var}(\hat{\sigma}^2) \! \\ 
	= \! \frac{\mathrm{Var}(\hat{\sigma}^2)}{4\sigma^2} \! = \! \frac{1}{4\sigma^2} \frac{(\kappa_d-1)\sigma^4}{t} \! = \! \frac{(\kappa_d-1)\sigma^2}{4t}.\end{gather}
Therefore the standard error (standard deviation of the estimated standard deviation) is
\begin{equation}SE_a(t)=\sqrt{\mathrm{Var}(g(\hat{\sigma}^2))} =\sqrt{\mathrm{Var}(\hat{\sigma})} = \frac{\sqrt{\kappa_d-1}}{2} \frac{\sigma}{\sqrt{t}} \propto \frac{\sigma}{\sqrt{t}}.\end{equation}

The value of $\kappa_d$ for a few distributions that we will use in the examples is given next. For the scaled Rademacher (Bernoulli) distribution of $x_a(t)$ we have $\kappa=1$, while $\kappa_d=2$ since $d(t)$ is a linear sum of two distributions. For $x_a(t)$ being a sum of two uniform distributions $\kappa=2.4$ and $\kappa_d=2.7$. For Gaussian distributions $\kappa=\kappa_d=3$. 

\begin{figure}[htbp]
	\centering
	\includegraphics[scale=.47]{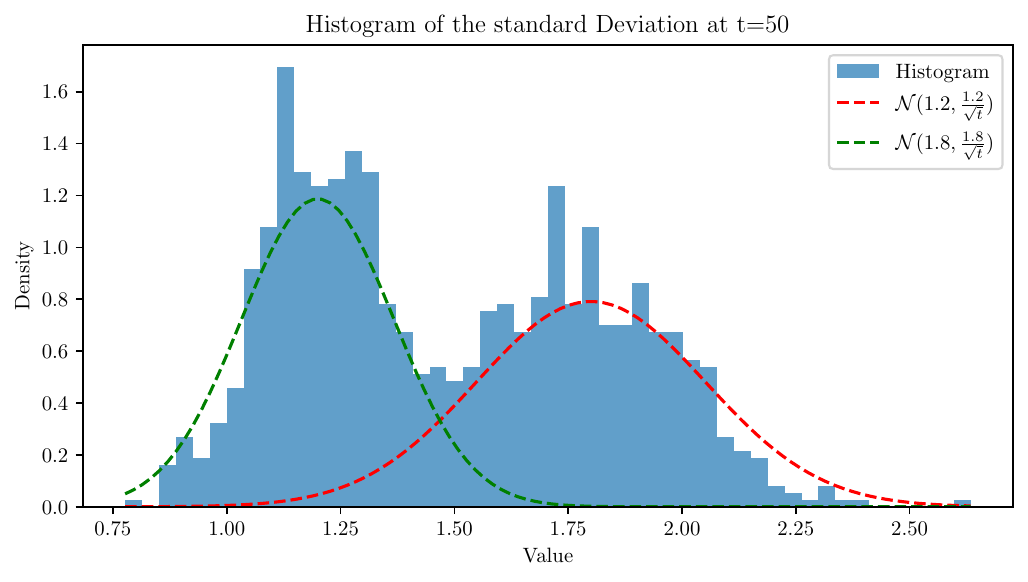}
	\includegraphics[scale=.47]{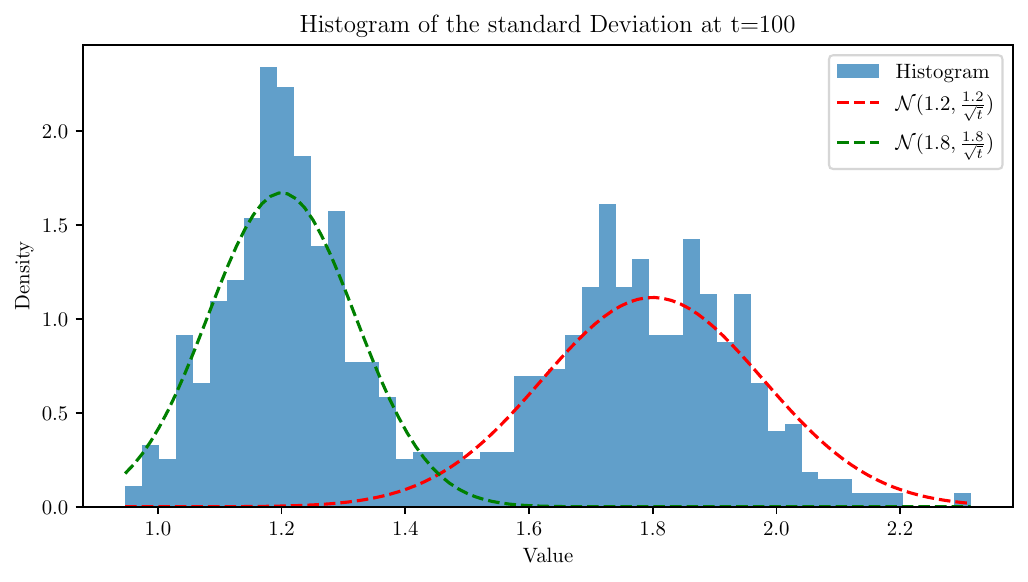}
	\includegraphics[scale=.47]{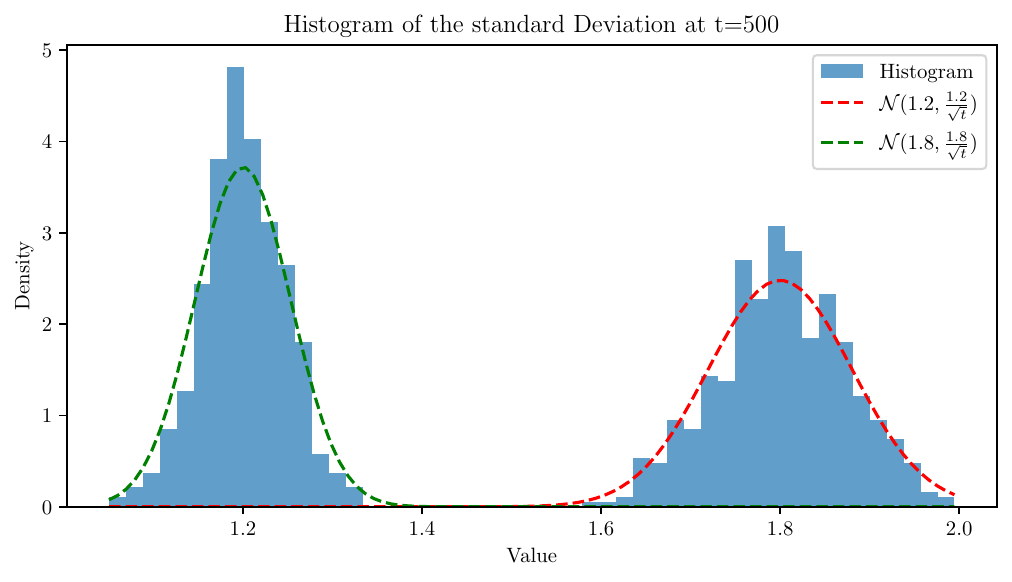}
	\includegraphics[scale=.47]{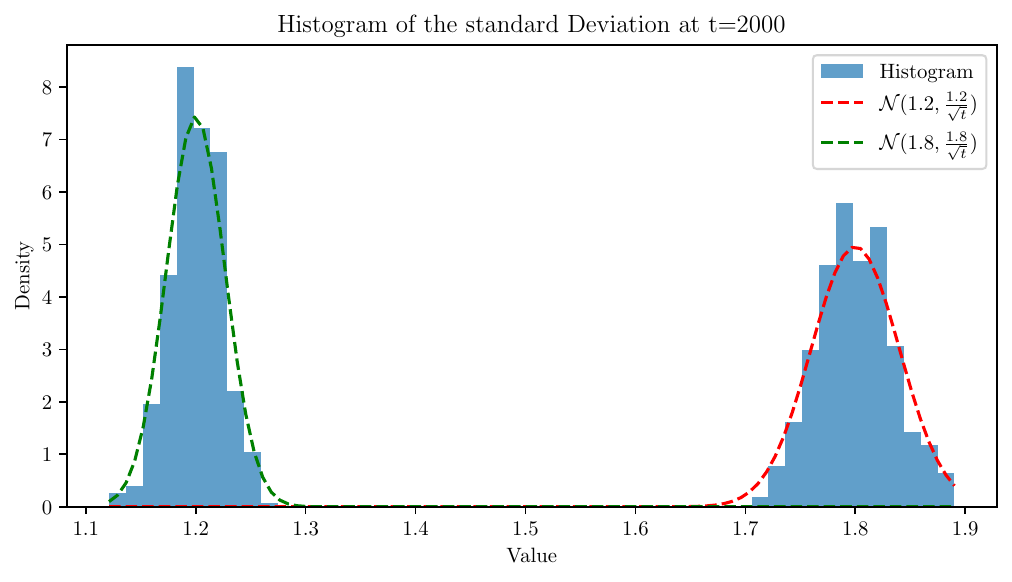}
	\caption{Histogram of the estimated sample standard deviation at different time instants $t$, converging to the derived theoretical distributions (dashed green and red line) as the number of data increases with $t$. High accuracy and agreement can be seen for $t=500$ and $t=2000$.}
	\label{Fig_5B_NW305_Hist}
\end{figure}

Since we used the first-order approximation in the Delta method, which tends to underestimate the variance, and $1/2 \le \sqrt{\kappa_d-1}/2 \le 1/\sqrt{2}$, for the considered distributions, we decided to further simplify the calculations that follow by using the same bounds for the standard error as for the local means.  This has been statistically checked comparing histograms with theoretically assumed distributions of standard deviations, Fig.  \ref{Fig_5B_NW305_Hist}.

Based on the previous assumptions and analysis, we can perform the confidence interval intersection check on both the local mean and the estimated standard deviation, defined by
\begin{gather}
	\mathbb{I}_a(t)= [ \bar{x}_{aa}(t) - \beta_{\delta}(t),  \ \ \bar{x}_{aa}(t) + \beta_{\delta}(t) ] \\
	\mathbb{I}^{\sigma}_a(t) = [ \hat{\sigma}_a(t) - \beta_{\delta}(t),  \ \ \hat{\sigma}_a(t) + \beta_{\delta}(t) ].
\end{gather}
Formally, we can consider this case as a two-dimensional data problem and use norm infinity, \cite{masterwithcode}, meaning that the graph edge between agents is disconnected if any of the confidence interval pairs $(\mathbb{I}_a(t), \ \ \mathbb{I}_{a'}(t))$ or $(\mathbb{I}^{\sigma}_a(t), \ \ \mathbb{I}^{\sigma}_{a'}(t))$ do not intersect. All other calculations (B-colME, C-colME) are the same as in the rest of the text. 

\noindent \textbf{Example:} Consider data with two similarity classes with very close means $(\mu_a, \mu_{a'})=(0.9, 1.1)$ and different standard deviations $(\sigma_a, \sigma_{a'})=(1.2, 1.8)$.  

The expected separation time for the local mean-based confidence intervals, according to the results of Section \ref{Sec:ExSep}, would occur very late, at 
\begin{equation}T_{sep}=4,265.\end{equation}

The expected separation time for standard deviation-based confidence intervals is obtained from $2\beta_{\delta}(t)=\sigma_a-\sigma_{a'}=1.8-1.2$ as
\begin{equation}T_{sep}=416.\end{equation}
Therefore the standard deviation based separation will dominate and drive the system. 

\begin{figure}[htbp]
	\centering
	
	\includegraphics[scale=0.8, trim={0 0 5.8cm 0}, clip]{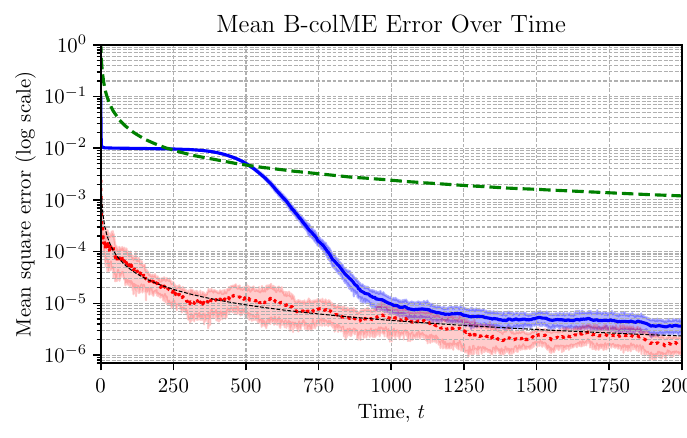}

	\includegraphics[scale=0.8, trim={0 0 5.3cm 0}, clip]{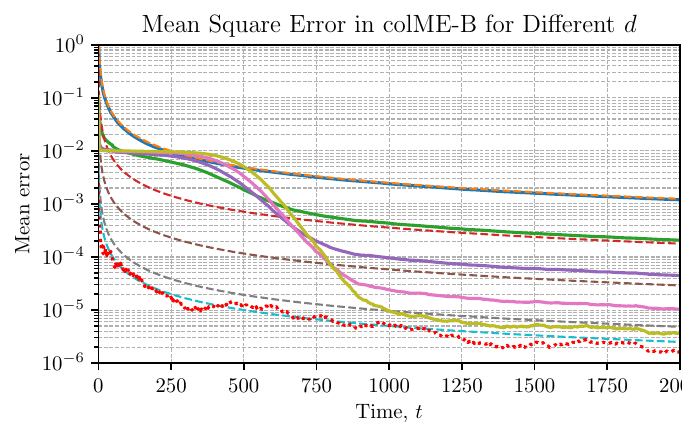}

	\caption{B-colME on data with the same mean, $N = 1000$, $\delta=0.01$. Total MSE of the
		estimation averaged over all agents at time instants $t$ in 10 realizations. In top panel: Local estimate (green dashed line); Proposed method (blue line) with corresponding bootstrap confidence intervals; Oracle solution  (red dotted line) with corresponding bootstrap confidence intervals. In bottom panel the MSE values for various depths $d=0, 1,2, 3, 4 $ are shown, with $d=0$ coinciding with local estimate and $d=4$ being close to oracle, while the others are in between in corresponding order. Theoretical lines are dashed.}
	\label{Fig_B225NW05_Rec_Var}
\end{figure}

\begin{figure}[htbp]
	\centering
	\includegraphics[scale=.8]{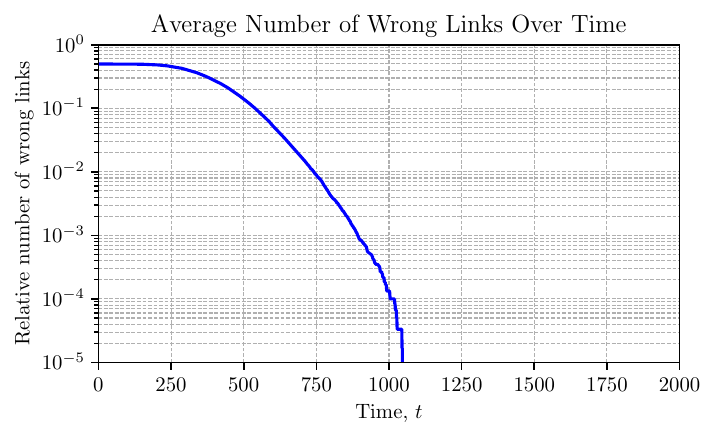}
	\caption{B-colME on data with the same mean: Relative total number of wrong links.}
	\label{Fig_5B_NW305_Var}
\end{figure}

For collaborative mean estimation, we used B-colME with $N=1000$ agents. The results are presented next, in Fig. \ref{Fig_5B_NW305_Hist}, Fig. \ref{Fig_B225NW05_Rec_Var}, and Fig. \ref{Fig_5B_NW305_Var}, for the standard deviation estimation, MSE, and wrong links over time. We can see that the region of the oracle solution MSE is reached at about $t=1000$ iterations.  While we checked both, intersection of the mean-based and standard deviation based confidence intervals, all disconnection of the wrong links have been done based of the standard deviation intervals, since their separation time is much lower.

\newpage

\begin{figure}[htbp]
	\centering
	
	\includegraphics[scale=0.8, trim={0 0 5.8cm 0}, clip]{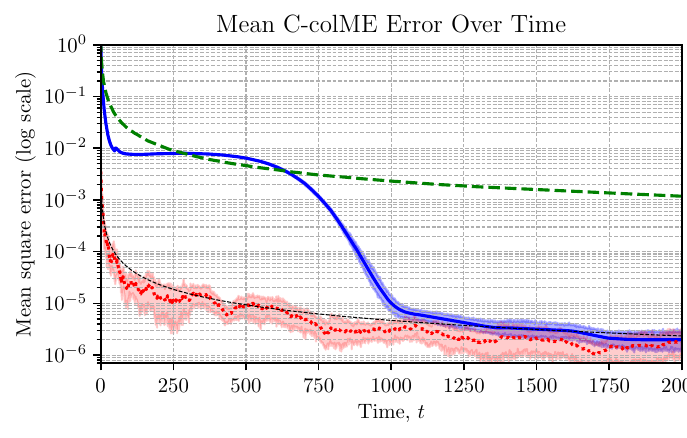}
	
	\caption{C-colME on data with similar mean, $N = 1000$, $\delta=0.01$. Total MSE of the
		estimation averaged over all agents in 10 realizations. Local estimate (green dashed line); Proposed method (blue line) with corresponding bootstrap confidence intervals; Oracle solution  (red dotted line).}
	\label{Fig_C225NW05_Rec_Var}
\end{figure}

\begin{figure}[htbp]
	\centering
	\includegraphics[scale=.8]{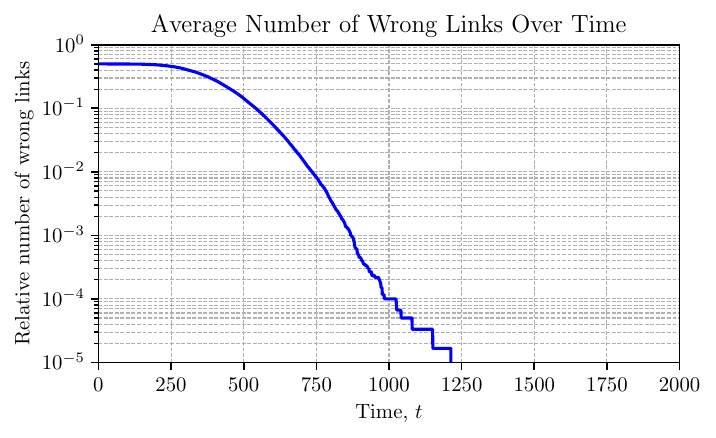}
	\caption{C-colME on data with similar mean: Relative total number of wrong links.}
	\label{Fig_5C_NW305_Var}
\end{figure}

\noindent \textbf{Example:}  We also considered the case with very close means (0.9,1.1) and different standard deviations (1.2, 1.8) with C-colME instead of B-colME and $N=1000$ agents. The results are shown in Figs. \ref{Fig_C225NW05_Rec_Var} and \ref{Fig_5C_NW305_Var}. The MSE approaches the oracle value just around $t=1000$, when the number of wrong links is very low.  All disconnections where based on the standard deviation confidence intervals.

\section{Three-Fold Confidence Intervals for Kurtosis, Standard Deviation, and Mean}

Now consider even more complex cases, where \textbf{both the true means and true standard deviations are very close, but the similarity class differ in distribution type}. For example, one is Gaussian distributed (with kurtosis equal to 3) and the other is uniformly distributed (with kurtosis equal to 1.8).  Here, we can use the confidence intervals based on the estimated kurtosis, 
\begin{equation}\hat{\kappa}_a(t) = \frac{1}{2t \hat{\sigma}^4_a(t)}\sum_{\tau=1}^t\big(x_{a}(\tau) - x_{a}(\tau-1)\big)^4-3,
\end{equation}
in addition to the confidence intervals based on the mean and the standard deviation. 

The statistics of kurtosis for the sub-Gaussian data is quite complex. However, it can be shown when the number of instants is sufficiently large (what is the case in our region of interest, when $t$ includes hundreds of samples) for the variance of kurtosis,  we can use Fisher approximation \cite{fisher1930moments}
\begin{equation}\sigma^2_{a,kurtosis}(t)=\frac{24t(t-1)^2}{(t-3)(t-2)(t+3)(t+5)} \approx \frac{24}{t}.\end{equation}
Then we can perform simultaneous confidence intervals check on all three parameters: the local mean, the estimated standard deviation, and the estimated kurtosis, using 
\begin{gather}
	\mathbb{I}_a(t)= [ \bar{x}_{aa}(t) - \beta_{\delta}(t),  \ \ \bar{x}_{aa}(t) + \beta_{\delta}(t) ] \\
	\mathbb{I}^{\sigma}_a(t) = [ \hat{\sigma}_a(t) - \beta_{\delta}(t),  \ \ \hat{\sigma}_a(t) + \beta_{\delta}(t) ]\\
	\mathbb{I}^{\kappa}_a(t) = [ \hat{\kappa}_a(t) - z_{\delta}  \frac{\sqrt{24}}{\sqrt{t}},  \ \ \hat{\kappa}_a(t) + z_{\delta} \frac{\sqrt{24}}{\sqrt{t}} ].
\end{gather}

For kurtosis we used confidence intervals defined under Gaussian assumption, but with very high probability $\delta$, resulting in wider intervals, since this assumption is rough. The histogram of kurtosis is always plotted along with its assumed distribution.

As an example, consider the expected separation time of the confidence intervals based on kurtosis for the Bernoulli distribution (expected kurtosis equal to 1) and the sum of 4 uniform random data (expected kurtosis 2.7) with confidence probability $\delta =0.001$ and $z_{\delta}=3.89$. From the relation $k_{4U}-k_B = 2 z_{\delta} \sqrt{24/t}$ in the form  \begin{equation} 2.7-1.0 = 2 \times 3.89 \sqrt{24/t}\end{equation}
we get $T_{sep}=502$. 

The problem in which the local mean, the local standard deviation, and the local kurtosis are used can be considered as a three-dimensional data problem, with each agent associated with
\begin{equation}\mathbf{x}_a(t) = \Big(\bar{x}_{aa}(t), \ \  \hat{\sigma}_{a}(t),  \ \   \hat{\kappa}_{a}(t)\Big).\end{equation}
Within this framework,  we will use the norm infinity, which means that the edge of the graph is disconnected if any pair of confidence intervals: \begin{equation}(\mathbb{I}_a(t), \ \mathbb{I}_{a'}(t)) \text{ or } (\mathbb{I}^{\sigma}_a(t),  \ \mathbb{I}^{\sigma}_{a'}(t))\text{ or }  (\mathbb{I}^{\kappa}_a(t), \ \mathbb{I}^{\kappa}_{a'}(t))\end{equation} does not intersect. 

\noindent \textbf{Example:} Consider the case with close means (0.9, 1.1), close standard deviations (1.9, 2.1), but with different distribution types.  One class was Bernoulli distributed with kurtosis equal to 1 and the other class was a sum of 4 uniformly distributed random variables with kurtosis equal to 2.7 (close to Gaussian when kurtosis is 3). 

In this case the confidence intervals for the mean and standard deviation are checked all the time from $t=0$ to $t=2000$. The expected separation time for both of them is obtained from $2\beta_{\delta}(t)=\sigma_a-\sigma_{a'}=\mu_a-\mu_{a'}$, as $T_{sep}=7,825$.  Since the kurtosis is calculated using the fourth power of data, we have waited and applied the kurtosis confidence interval checks after $t=500$, with reconnection option being active \cite{masterwithcode}.  We have checked and concluded that all edge disconnections in this case were based on the kurtosis intervals, since the mean and the standard deviation are too close and their separation time would come much later. The histograms of kurtosis for various values of $t$ are given next.

\begin{figure}[htbp]
	\centering
	
	\includegraphics[scale=0.8, trim={0 0 5.8cm 0}, clip]{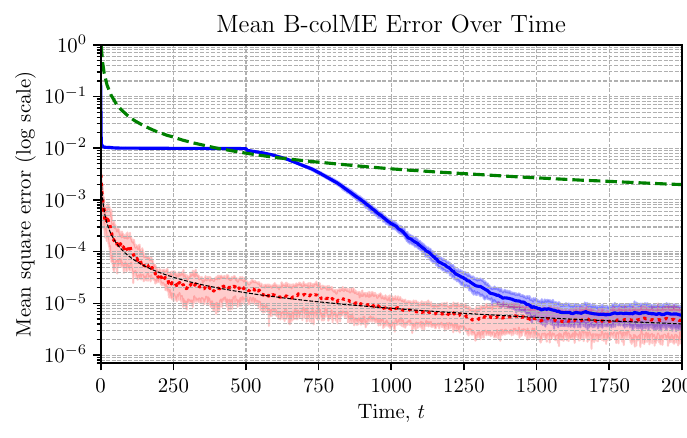}

	\includegraphics[scale=0.8, trim={0 0 5.3cm 0}, clip]{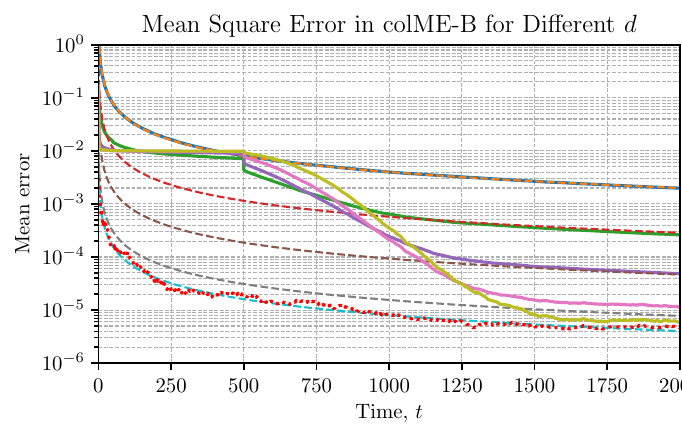}

	\caption{
		B-colME on data with the same mean, $N = 1000$, $\delta=0.01$. Total MSE of the
		estimation averaged over all agents at time instants $t$ in 10 realizations. In top panel: Local estimate (green dashed line); Proposed method (blue line) with corresponding bootstrap confidence intervals; Oracle solution  (red dotted line) with corresponding bootstrap confidence intervals. In bottom panel the MSE values for various depths $d=0, 1,2, 3, 4 $ are shown, with $d=0$ coinciding with local estimate and $d=4$ being close to oracle, while the others are in between in a corresponding order. Theoretical solution lines are dashed.
		}
	\label{Fig_B225NW05_Rec_kur}
\end{figure}

\begin{figure}[htbp]
	\centering
	\includegraphics[scale=.8]{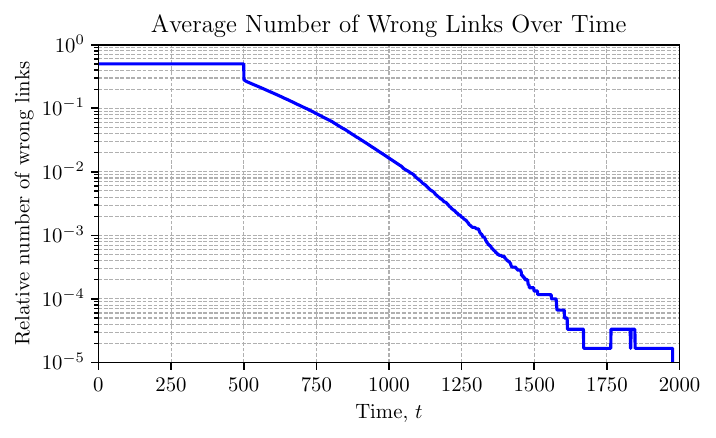}
	\caption{B-colME on data with similar mean and variance, different distribution of data: Relative total number of wrong links. Since the kurtosis is calculated using the fourth power of data, we have waited and applied the kurtosis confidence interval checks after $t=500$, with reconnection option being active.}
	\label{Fig_5B_NW305_Kur}
\end{figure}

The collaboration results with B-colME are given in Fig. \ref{Fig_B225NW05_Rec_kur} and Fig. \ref{Fig_5B_NW305_Kur}.

\begin{figure}[htbp]
	\centering
	\includegraphics[scale=.55]{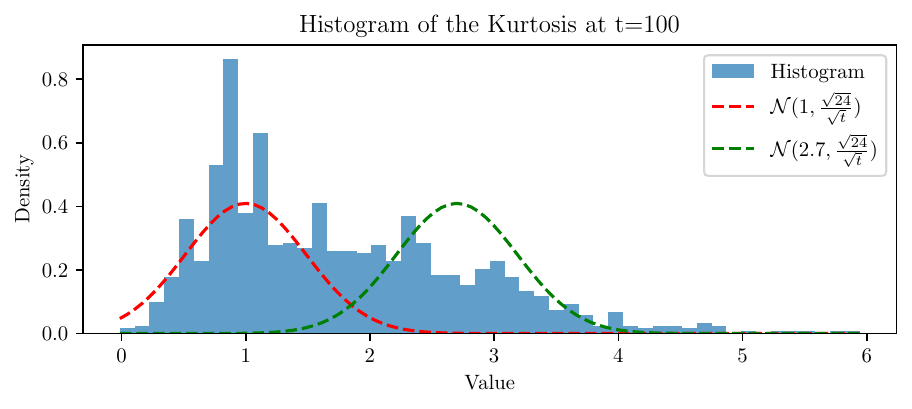}
	\includegraphics[scale=.55]{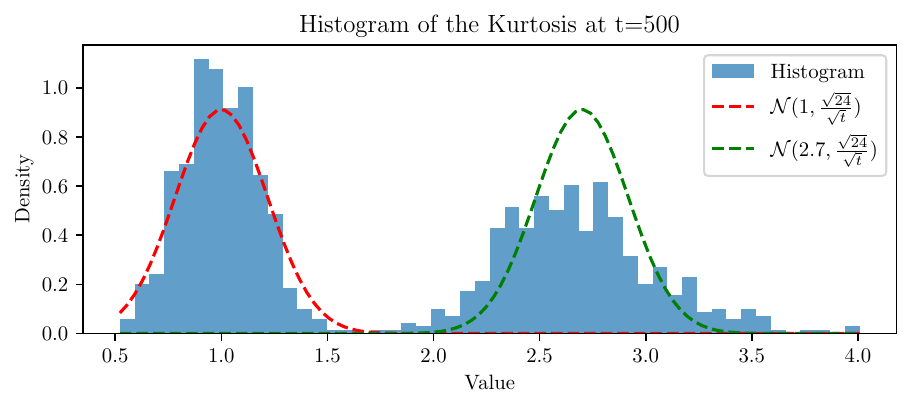}
	\includegraphics[scale=.55]{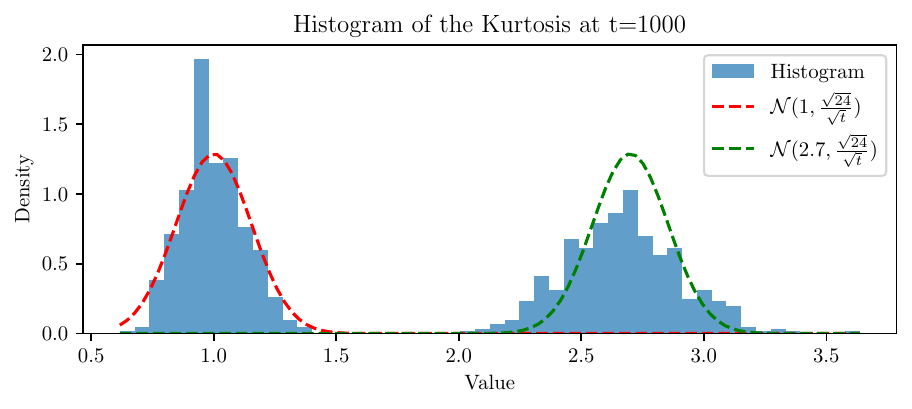}
	\includegraphics[scale=.55]{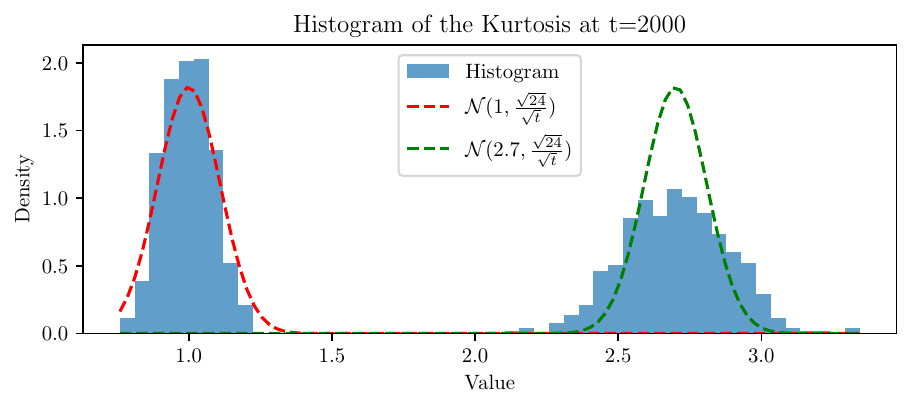}
	\caption{Histogram of the estimated sample kurtosis  at diffrent time instants $t$, converging to the derived theoretical distributions (dashed green and red line) as the number of data increases with $t$.}
	\label{Fig_5B_NW305_Hist_kur}
\end{figure}    

Figure~\ref{Fig_5B_NW305_Hist_kur} shows histograms of the estimated kurtosis for agents at different time instants, along with the corresponding theoretical distributions. Initially, the distributions are overlapping, but over time the histograms reflect the differences in underlying data distributions, highlighting class separation based on kurtosis.

Figure~\ref{Fig_B225NW05_Rec_kur} presents the total mean squared error (MSE) over time for B-colME applied to data with similar mean and variance but different distributions. The top panel shows the MSE for a fixed neighborhood, while the bottom panel shows the MSE for various neighborhood sizes. The decreasing MSE indicates improved estimation as agents segregate according to distributional differences.

Figure~\ref{Fig_5B_NW305_Kur} shows the total number of wrong links over time. Despite differences in distribution, the number of incorrect inter-class connections behaves consistently, reflecting the effect of confidence interval-based edge pruning on the network structure.

\section{Multiclass Example with Kurtosis, Standard Deviation, and Mean Confidence Intervals}

Consider now a case with, for example, four classes of agents and no single confidence interval approach can be used. For some pairs of classes, the means are so close that they cannot be separated using local means, but variances may be different, while for other pairs of classes the means are distant, but variances are close, while there are pairs of classes when only the distribution type significantly differs one class from another. 

In specific, consider the following classes, with corresponding means, standard deviations, distribution types:
\begin{equation}\text{\textbf{Class 1}: }(\mu_1, \ \sigma_1)=(0.9, \ \ 1.8) \ \ \ \text{sum of two uniform distributions}\nonumber \end{equation}
\begin{equation}\text{\textbf{Class 2}: } (\mu_2, \ \sigma_2)=(0.1, \ \ 2.0)  \ \ \ \text{sum of two uniform distributions}\nonumber \end{equation}
\begin{equation}\text{\textbf{Class 3}: } (\mu_3, \ \sigma_3)=(1.1, \ \ 1.2) \ \ \ \text{sum of two uniform distributions} \nonumber \end{equation}
\begin{equation}\text{\textbf{Class 4}: } (\mu_4, \ \sigma_4)=(1.0, \ \ 1.9) \ \ \ \text{Bernoulli distribution}\end{equation}

\begin{table}[ht]
	\centering
	\caption{Table with colored symbols indicating what confidence intervals will be used for separation of agents: Intervals $(\mathbb{I}_a(t), \ \mathbb{I}_{a'}(t))$,  $(\mathbb{I}^{\sigma}_a(t),  \ \mathbb{I}^{\sigma}_{a'}(t))$, and $(\mathbb{I}^{\kappa}_a(t), \ \mathbb{I}^{\kappa}_{a'}(t))$ are given in this respective order.  Red circle means that this form is not used for separation (expected separation time is very late), while green means that this interval can be used (reasonable separation time). Of course, in all cases all confidence intervals are checked.}
	
	Agent classes (means, standard deviations, kurtosis) intervals

	\begin{tabular}{|c|c|c|c|c|}
		\hline
		& 1 & 2 & 3 & 4 \\
		
		\hline
		1 &  & \textcolor{green}{\scalebox{2.0}{$\bullet$}} \textcolor{red}{\scalebox{2.0}{$\bullet$}} \textcolor{red}{\scalebox{2.0}{$\bullet$}} & \textcolor{red}{\scalebox{2.0}{$\bullet$}} \textcolor{green}{\scalebox{2.0}{$\bullet$}} \textcolor{red}{\scalebox{2.0}{$\bullet$}} & \textcolor{red}{\scalebox{2.0}{$\bullet$}} \textcolor{red}{\scalebox{2.0}{$\bullet$}} \textcolor{green}{\scalebox{2.0}{$\bullet$}}  \\ 
		& & $({\color{green}373}, {\color{red}7023, \infty})$ & $({\color{red}4264}, {\color{green}416}, {\color{red}\infty} )$ & $({\color{red}28552, 28552}, {\color{green}741}) $ \\
		\hline
		2 & \textcolor{green}{\scalebox{2.0}{$\bullet$}} \textcolor{red}{\scalebox{2.0}{$\bullet$}} \textcolor{red}{\scalebox{2.0}{$\bullet$}} &  & \textcolor{green}{\scalebox{2.0}{$\bullet$}} \textcolor{green}{\scalebox{2.0}{$\bullet$}} \textcolor{red}{\scalebox{2.0}{$\bullet$}} & \textcolor{green}{\scalebox{2.0}{$\bullet$}} \textcolor{red}{\scalebox{2.0}{$\bullet$}} \textcolor{green}{\scalebox{2.0}{$\bullet$}} \\ 
		& &  & $({\color{green}161}, 258, {\color{red}\infty} )$ & $({\color{green}306}, {\color{red}31890}, 741) $ \\
		\hline
		3 & \textcolor{red}{\scalebox{2.0}{$\bullet$}} \textcolor{green}{\scalebox{2.0}{$\bullet$}} \textcolor{red}{\scalebox{2.0}{$\bullet$}} & \textcolor{green}{\scalebox{2.0}{$\bullet$}} \textcolor{green}{\scalebox{2.0}{$\bullet$}} \textcolor{red}{\scalebox{2.0}{$\bullet$}} &  & \textcolor{red}{\scalebox{2.0}{$\bullet$}} \textcolor{green}{\scalebox{2.0}{$\bullet$}} \textcolor{green}{\scalebox{2.0}{$\bullet$}} \\ 
		& &  &  & $({\color{red}19685},{\color{green}321}, 741) $ \\
		\hline
		4 & \textcolor{red}{\scalebox{2.0}{$\bullet$}} \textcolor{red}{\scalebox{2.0}{$\bullet$}} \textcolor{green}{\scalebox{2.0}{$\bullet$}} & \textcolor{green}{\scalebox{2.0}{$\bullet$}} \textcolor{red}{\scalebox{2.0}{$\bullet$}} \textcolor{green}{\scalebox{2.0}{$\bullet$}} & \textcolor{red}{\scalebox{2.0}{$\bullet$}} \textcolor{green}{\scalebox{2.0}{$\bullet$}} \textcolor{green}{\scalebox{2.0}{$\bullet$}} &  \\ \hline
	\end{tabular}
	\label{Tab:Ex_sep_kur}
\end{table}

Expected separation times for the local mean based confidence intervals, standard deviation confidence intervals, and kurtosis based confidence intervals are calculated by solving for $t=T_{sep}$ the following equalities:
\begin{gather}
	2\beta_{\delta}(t)=\mu_a-\mu_{a'}\\
	2\beta_{\delta}(t)=\sigma_a-\sigma_{a'}\\
	2 z_{\delta} \sqrt{\frac{24}{t}}=\kappa_a-\kappa_{a'}.
\end{gather}
They are given in Table \ref{Tab:Ex_sep_kur}, as triples corresponding to the mean, standard deviation, and kurtosis, respectively. 

\bigskip

\noindent\textbf{Approximate Class Separation Analysis:}
\begin{enumerate}
	\item
	The expected separation time for the \textit{confidence intervals based on kurtosis} (for a sum of two uniform distributions with kurtosis equal to 2.4 and the Bernoulli distribution with kurtosis equal to 1) is obtained from
	\begin{equation}2.4-1.0=2 \times 3.89 \sqrt{24/t}\end{equation}
	as $T_{sep}=741$. The statistical value can be slightly higher since we will not allow disconnection according to this criterion before the local kurtosis is sufficiently accumulated (for example, until $t=500$).

	\item The expected separation time based on \textit{the standard deviation confidence intervals} between the first and third classes will be approximately $T_{sep}=416$ and between the third and fourth is $T_{sep}=321$ with an average of 369.  
	
	\item The expected separation time based on \textit{the local mean confidence intervals} for the first and second classes will be $T_{sep}=373$, for the second and third class at $T_{sep}=161$, and between the second and fourth $T_{sep}=306$, with an average of 280. 
	
	\item Statistical mean separation times slightly differ from the theoretical expected separation times. In our example, the separation based on the local mean starts first with no other concurrent criteria separating edges earlier. Later, separation based on the variance can take some of the disconnections for class two and three, making a lower number of mean-based disconnections for higher times, resulting in a slightly lower statistical average. Since kurtosis-based disconnections start last, some of its edges are earlier disconnected by mean or standard deviation, making its average separation time slightly longer. 
\end{enumerate}

\begin{figure}[ht]
	\centering
	\includegraphics[scale=.8]{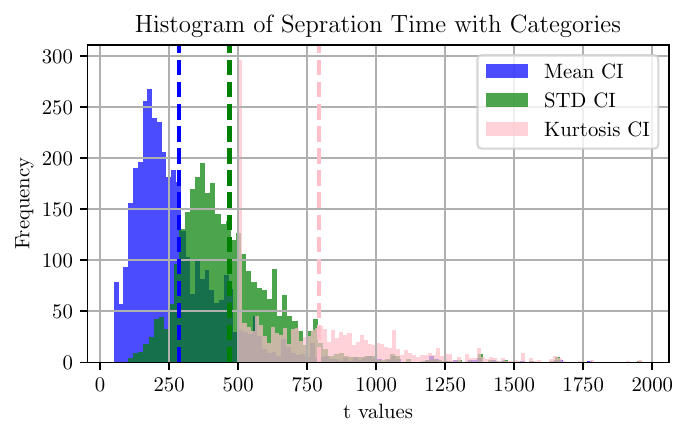}
	
	\caption{Histogram of separation times with four classes of agents and no single confidence interval approach can be used. Mean separation times are given by the vertical dashed lines. They are close to the theoretical expected separation times for the given classes of data. }
	\label{Fig_B225NW05_Rec_kur_4C_H_Tsep}
\end{figure}

\begin{figure}
		
		\centering
		
		\includegraphics[scale=.35]{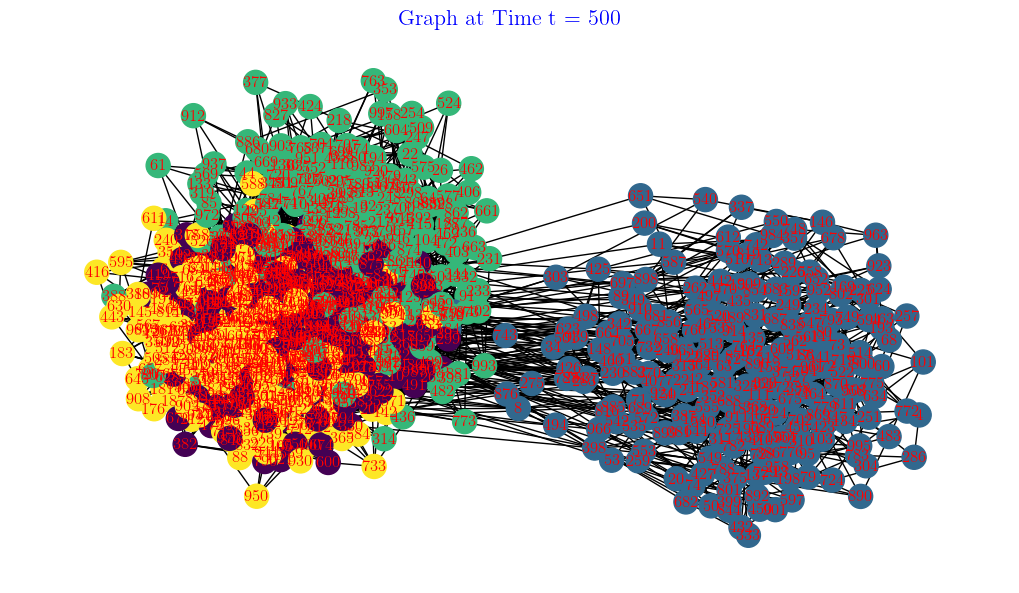}
		
		\includegraphics[scale=.35]{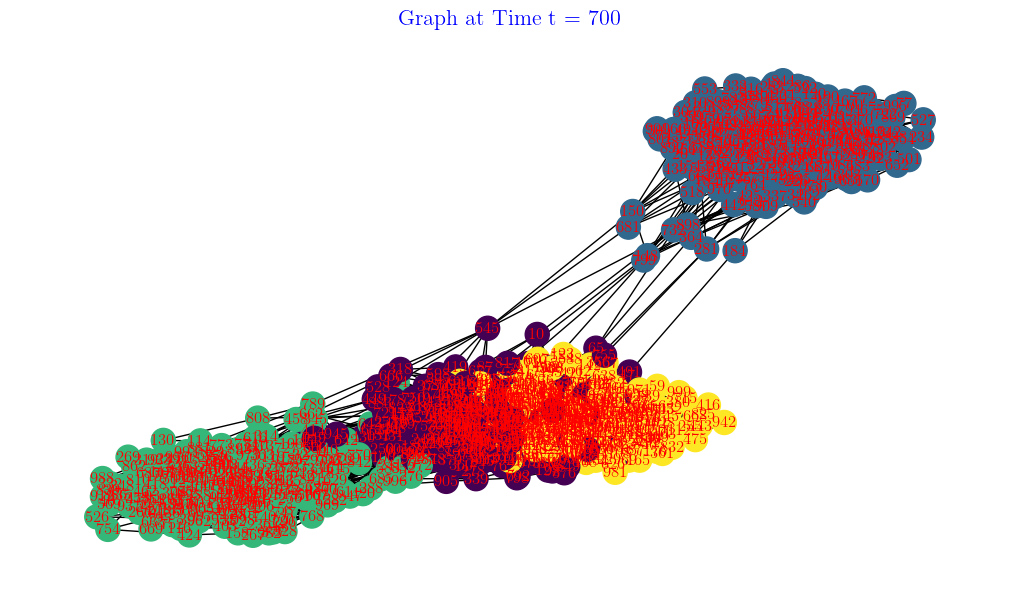}
		
		\includegraphics[scale=.35]{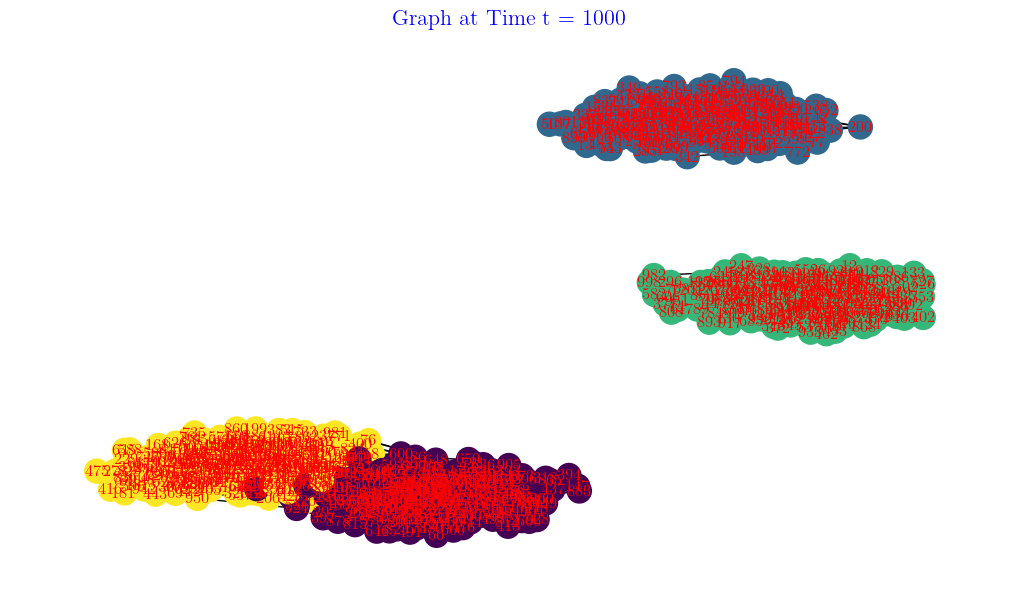}
		
		\includegraphics[scale=.35]{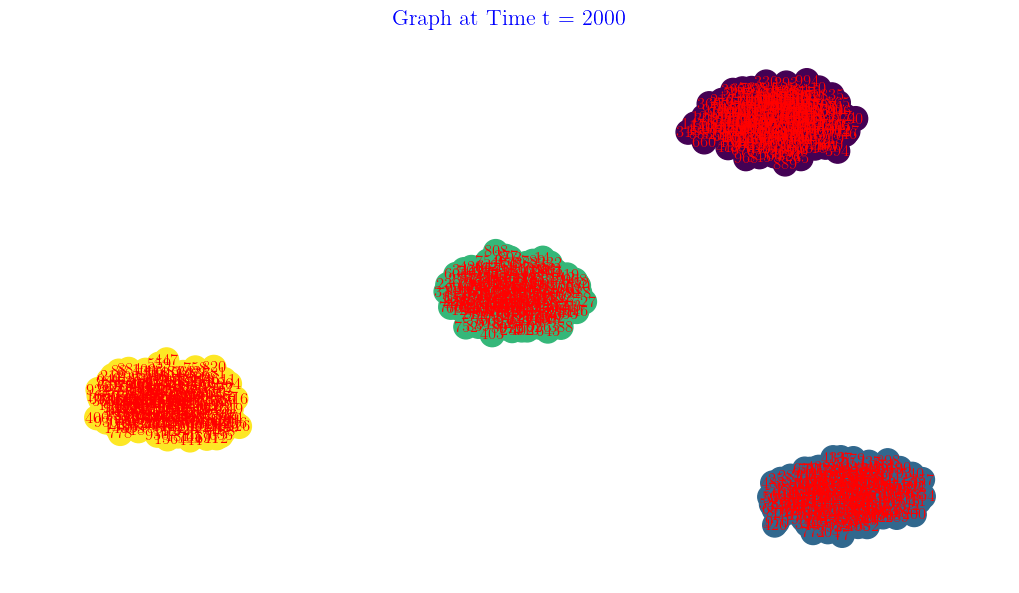}

	\caption{B-colME of  data with $N=1,000$ with four classes of agents and no single confidence interval approach can be used. Connection graphs at $t=500$, $t=700$, $t=1000$,  and $t=2000$, respectively, are shown from top to bottom.}
	\label{Fig_1C_3D_MV_kur}
\end{figure}

The separation times are recorded in the program and their histograms are shown in Fig. \ref{Fig_B225NW05_Rec_kur_4C_H_Tsep}, where the mean value of the recorded separation time is also calculated and shown for each type of confidence intervals by vertical dashed lines. They are in agreement with the above approximate theoretical analysis. From Table \ref{Tab:Ex_sep_kur} we can see that class 2 will separate first (its highest expected separation time is 373), then class 3 will separate (with the highest expected separation time 416) and finally classes 1 and 4 will separate with the highest expected separation time of 741 (see Fig. \ref{Fig_1C_3D_MV_kur}).

\begin{figure}[htbp]
	\centering
	\includegraphics[scale=0.8]{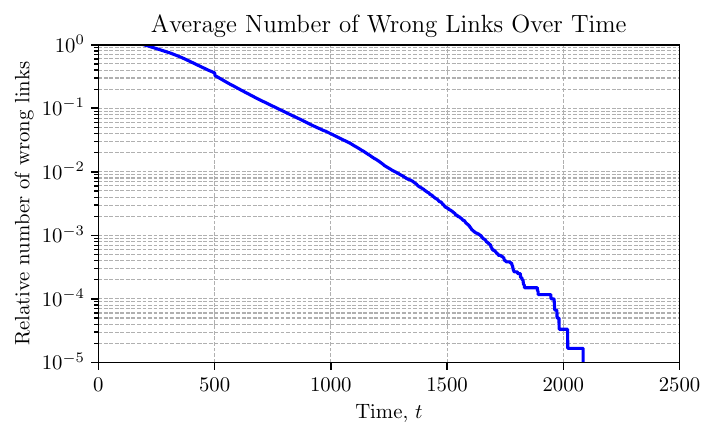}
	\caption{B-colME of  data with $N=1,000$ with four classes of agents and no single confidence interval approach can be used: Relative total number of wrong links.}
	\label{Fig_5B_NW305_Kur_4C}
\end{figure}

Here we used a B-colME with $N=1000$ agents and the mean square error and the number of wrong links over time are shown, along with graphs at a few representative instants. Figs. \ref{Fig_1C_3D_MV_kur}, \ref{Fig_1C_3D_MV_kur}. and \ref{Fig_5B_NW305_Kur_4C}. 

\begin{figure}[htbp]
	\centering
	
	\includegraphics[scale=0.8, trim={0 0 5.8cm 0}, clip]{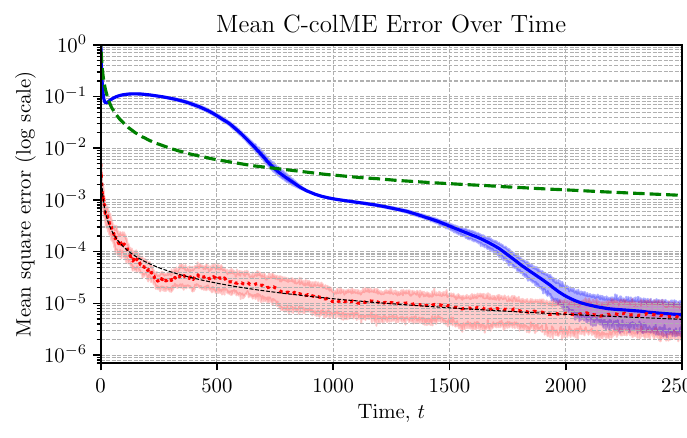}
	
	\caption{C-colME of  data with $N=1,000$ with four classes of agents and no single confidence interval approach can be used. Total MSE of the
		estimation averaged over all agents at time instants $t$ in 10 realizations.}
	\label{Fig_B225NW05_Rec_kur_4C_11}
\end{figure}

We have also implemented C-colME with agents $N=1000$ and the mean square error. Fig. \ref{Fig_B225NW05_Rec_kur_4C_11}.

\medskip

\noindent \textbf{Weighted graph:} Finally, the weighted graph is used to improve the convergence on both B-colME and C-colME, as introduced in \cite{masterwithcode}. For weights we used the smallest of three weights, corresponding to the least confidence intervals intersection. We calculated  \begin{equation}aW_{a,a'}(t)=e^{- \Big(2 \frac{\bar{x}_{a,a}(t)-\bar{x}_{a',a'}(t)}{2\beta_{\delta}(t)}\Big)^4} \text{ when } A_{a,a'}(t)=1.\end{equation} Then we calculated $e^{- \Big(2 \frac{\hat{\sigma}_{a}(t)-\hat{\sigma}_{a'}(t)}{2\beta_{\delta}(t)}\Big)^4}$ and used this value for $AW_{a,a'}(t)$ if it is smaller than the existing $AW_{a,a'}(t)$,
\begin{equation}aW_{a,a'}(t)=\min \{AW_{a,a'}(t), e^{- \Big(2 \frac{\hat{\sigma}_{a}(t)-\hat{\sigma}_{a'}(t)}{2\beta_{\delta}(t)}\Big)^4}\}\end{equation}
Finally, for $A_{a,a'}(t)=1$ we calculated  $v(t)=e^{- \Big(2 \frac{\hat{\kappa}_{a}(t)-\hat{\kappa}_{a'}(t)}{2z_{\delta}\sqrt{24/t}}\Big)^4}$,  compared it with the existing $AW_{a,a'}(t)$ and used 
\begin{equation}aW_{a,a'}(t)=\min \{ AW_{a,a'}(t),e^{- \Big(2 \frac{\hat{\kappa}_{a}(t)-\hat{\kappa}_{a'}(t)}{2z_{\delta}\sqrt{24/t}}\Big)^4} \}.\end{equation}

\begin{figure}[htbp]
	\centering
	
	\includegraphics[scale=0.8, trim={0 0 5.8cm 0}, clip]{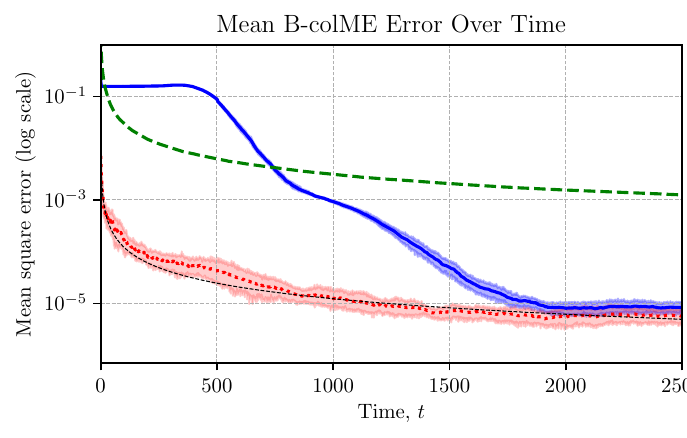}

	\includegraphics[scale=0.8, trim={0 0 5.3cm 0}, clip]{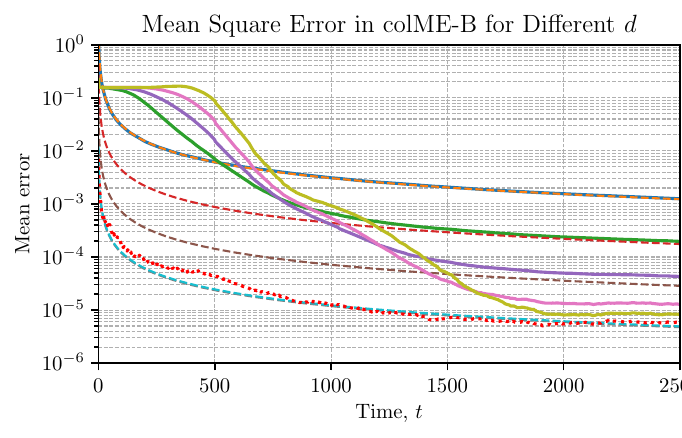}

	\caption{B-colME on data with the same mean, $N = 1000$, $\delta=0.01$. Total MSE of the
		estimation averaged over all agents at time instants $t$ in 10 realizations. In top panel: Local estimate (green dashed line); Proposed method (blue line) with corresponding bootstrap confidence intervals; Oracle solution  (red dotted line) with corresponding bootstrap confidence intervals. In bottom panel the MSE values for various depths $d=0, 1,2, 3, 4 $ are shown, with $d=0$ coinciding with local estimate and $d=4$ being close to oracle, while the others are in between in corresponding order. Theoretical solution  lines are dashed.}
	\label{Fig_B225NW05_Rec_kur_4C}
\end{figure}

\begin{figure}[htbp]
	\centering
	
	\includegraphics[scale=0.8, trim={0 0 5.8cm 0}, clip]{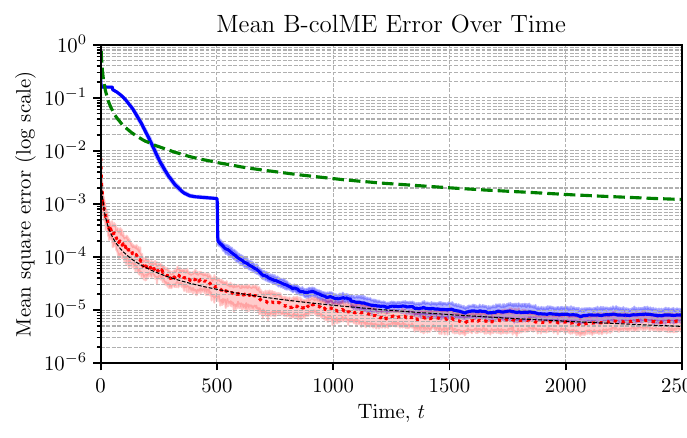}

	\includegraphics[scale=0.8, trim={0 0 5.3cm 0}, clip]{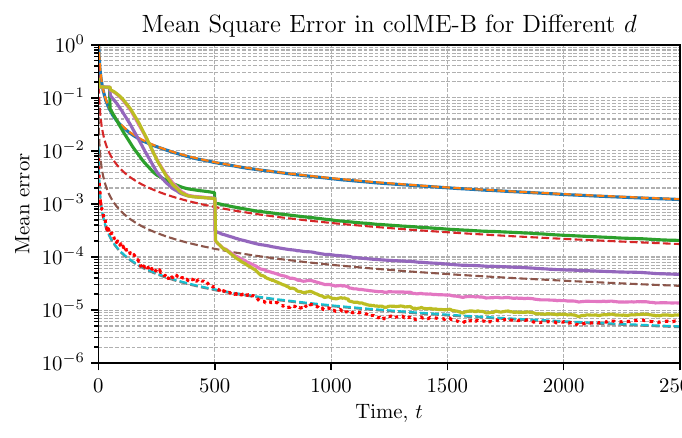}

	\caption{
		B-colME on data with the same mean, $N = 1000$, $\delta=0.01$,  with four classes of agents and no single confidence interval approach can be used. Total MSE of the
		estimation averaged over all agents at time instants $t$ in 10 realizations. In top panel: Local estimate (green dashed line); Proposed method (blue line) with corresponding bootstrap confidence intervals; Oracle solution  (red dotted line) with corresponding bootstrap confidence intervals. In bottom panel the MSE values for various depths $d=0, 1,2, 3, 4 $ are shown, with $d=0$ coinciding with local estimate and $d=4$ being close to oracle, while the others are in between in corresponding order. Theoretical lines are dashed.
		}
	\label{Fig_B225NW05_AW_Rec_kur_4C}
\end{figure}

\begin{figure}[htbp]
	\centering
	
	\includegraphics[scale=0.8, trim={0 0 5.8cm 0}, clip]{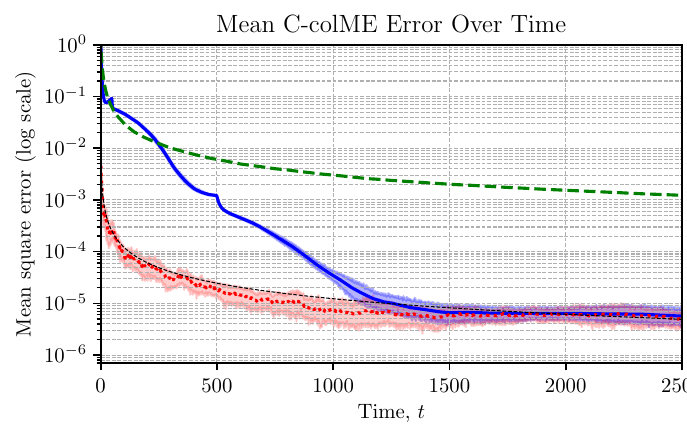}
	
	\caption{C-colME with weighted graph of  data with $N=1,000$ with four classes of agents and no single confidence interval approach can be used. Total MSE of the
		estimation averaged over all agents at time instants $t$ in 10 realizations. Since the kurtosis is calculated using the fourth power of data, we have waited and applied the kurtosis confidence interval checks after $t=500$, with reconnection option being active.}
	\label{Fig_B225NW05_Rec_kur_4C_1}
\end{figure}

The results, with an evident improvement in convergence compared to Figs. \ref{Fig_B225NW05_Rec_kur_4C} and \ref{Fig_B225NW05_Rec_kur_4C_11}, are shown in
Figs. \ref{Fig_B225NW05_AW_Rec_kur_4C} and \ref{Fig_B225NW05_Rec_kur_4C_1}.

\begin{figure}
	\centering
	\includegraphics[scale=.6]{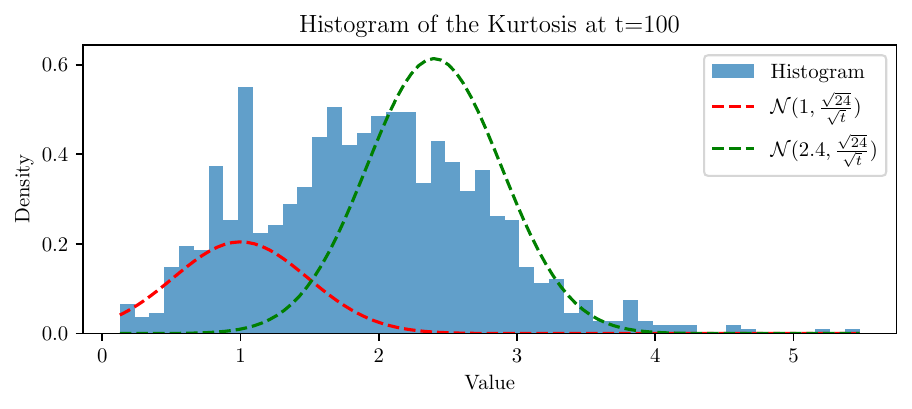}
	
	\includegraphics[scale=.6]{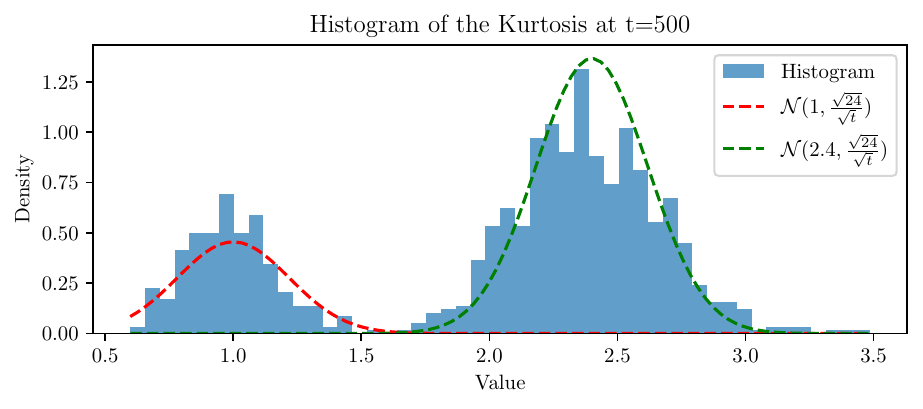}
	
	\includegraphics[scale=.6]{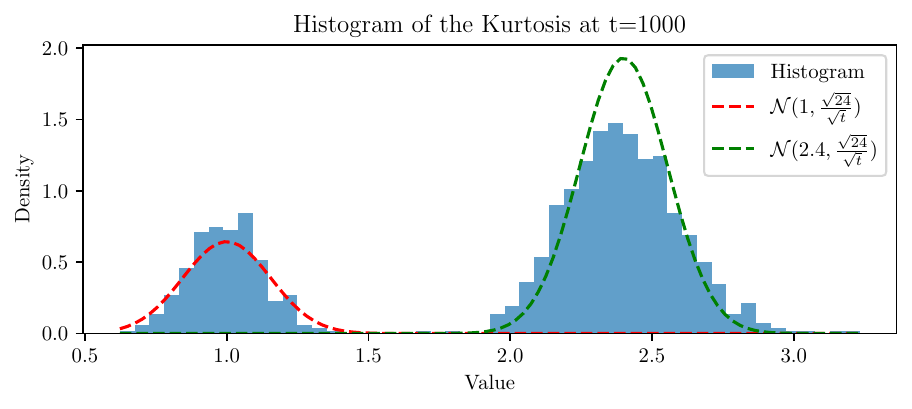}
	
	\includegraphics[scale=.6]{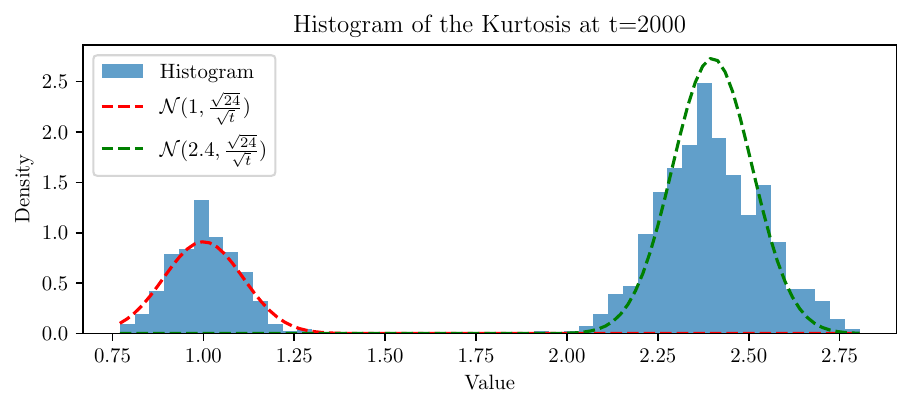}
	\caption{Histogram of the estimated sample kurtosis  at diffrent time instants $t$, converging to the derived theoretical distributions (dashed green and red line) as the number of data increases with $t$.}
	\label{Fig_5B_NW305_Hist_kur_4C}
\end{figure}

\begin{figure}
	\centering    
	
	\includegraphics[scale=.6]{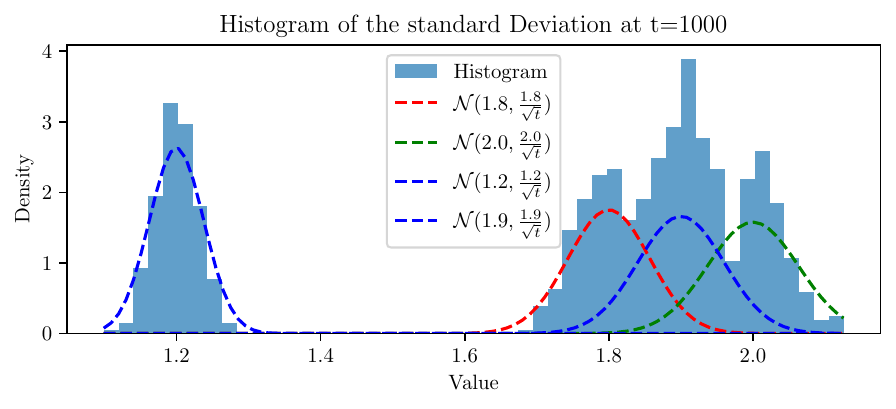}
	
	\includegraphics[scale=.6]{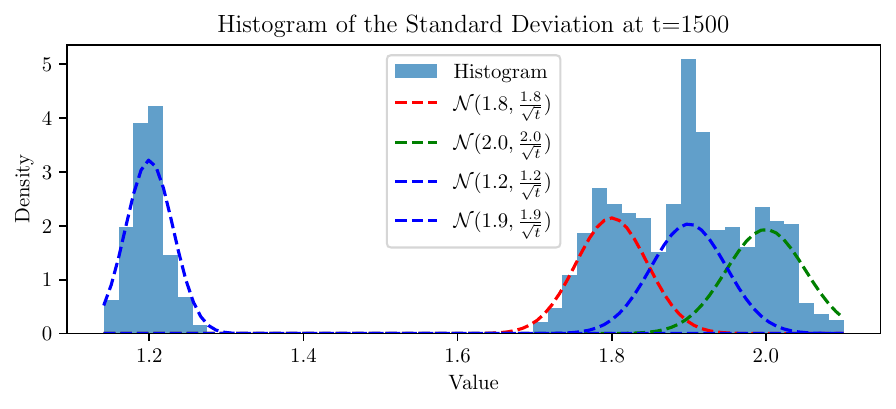}
	
	\includegraphics[scale=.6]{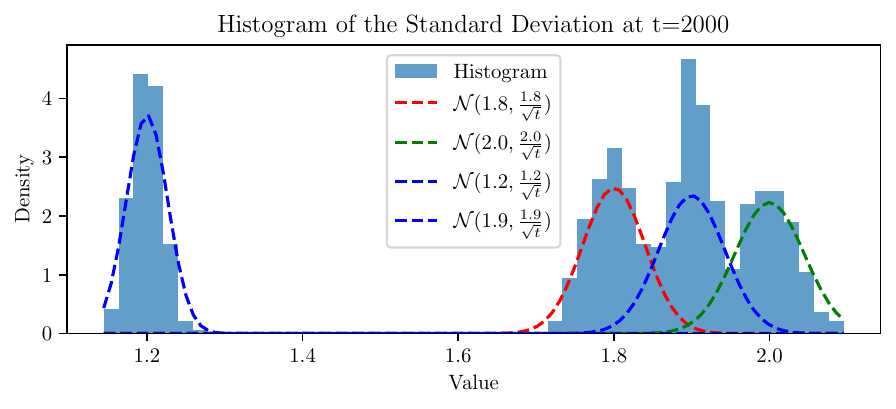}
	
	\includegraphics[scale=.6]{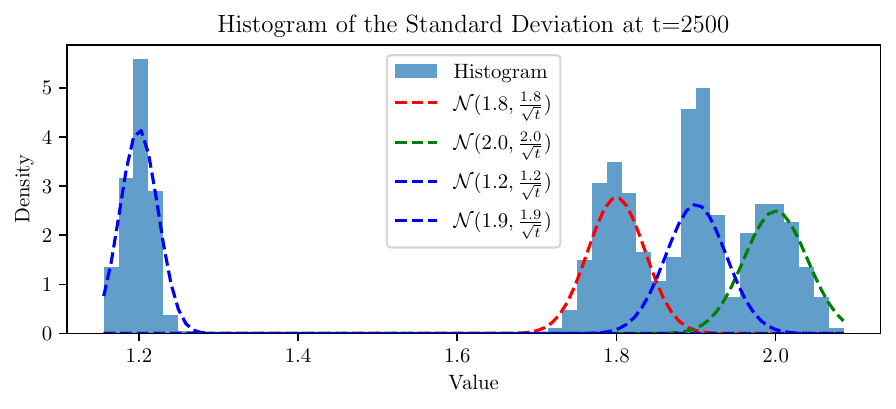}
	
	\caption{Histogram of the estimated sample standard deviation at different time instants $t$, converging to the derived theoretical distributions (dashed, blue, red, green, blue,  and green line) as the number of data increases with $t$.}
	\label{Fig_5B_NW305_Hist_var_4C}
	
\end{figure}

Figure~\ref{Fig_5B_NW305_Hist_kur_4C} shows histograms of the kurtosis for agents across four classes at different time instants. The distributions initially overlap but gradually reflect the differences among the four classes, indicating potential separation based on distribution.

Figure~\ref{Fig_5B_NW305_Hist_var_4C} presents histograms of the estimated standard deviations over time. Variability among classes becomes more distinct as agents evolve, highlighting the role of variance in class separation. Initially, the first class is separated from the others, while the other three are still seemingly merged together. After a while, $t=2000$, three distinct peaks can be observed, while still significantly overlapping, matching the theoretical distributions and corresponding to four classes at $t=2500$.

Figure~\ref{Fig_1C_3D_MV_kur} illustrates the evolution of the network graph. Initially, agents are arbitrarily connected. Over time, edges are adjusted based on mean, variance, and kurtosis confidence intervals, progressively separating the four classes, with full clustering of  four distinct classes by $t=2000$.

Figure~\ref{Fig_B225NW05_Rec_kur_4C} shows the total MSE over time for B-colME with four classes. The MSE decreases as agents segregate correctly, with different neighborhood sizes influencing the convergence rate. The MSE of the approach of the proposed algorithm is shown as the blue line, associated with its confidence intervals. To confirm convergence to the oracle solution, we can observe that blue line lies on the red dashed line at slightly after $t=1800$. As for B-colME, all approaches (at various depths) approach towards their theoretical solutions, with the fastest being a t $d=1$, with around $t=1000$ and slowest being at $d=4$. The B-colME with $d=4$ converges to a value close to the oracle MSE, slightly before $t=2000$

Figure~\ref{Fig_5B_NW305_Kur_4C} displays the total number of wrong links,  inter-class connections over time. As separation progresses, the number of wrong links diminishes, reflecting improved class distinction.

Figure~\ref{Fig_B225NW05_Rec_kur_4C_11} presents the total MSE for C-colME with four classes. The decrease in MSE demonstrates successful collaboration and accurate estimation despite complex class distributions. The convergence to the region of the oracle solution is reached at $t=2000$. This means that the all four classes, even in this complex case, will be separated after a certain time, using the combined intersection of the confidence intervals introduced.

Figures~\ref{Fig_B225NW05_AW_Rec_kur_4C} and \ref{Fig_B225NW05_Rec_kur_4C_1} show the MSE for B-colME and C-colME, respectively, on weighted graphs. Weighting improves convergence by emphasizing the most confident connections and decreasing the weight of the least confident connections,, resulting in  lower estimation errors. The convergence here is reached faster than the unweighted approaches, with oracle region being reached at around $t=1000$ for both B-colME and C-colME cases

 More examples and implementation details may be found in \cite{masterwithcode}.

\section{Conclusion}
. 
In this paper, we investigated the problem of collaborative mean estimation in decentralized and heterogeneous settings, with a particular focus on improving convergence behavior. Building upon the ColME framework, we examined how confidence interval design and graph-based collaboration affect the ability of agents to identify suitable collaborators and efficiently aggregate information. This study presents a decentralized framework for the online, local, and collaborative estimation of sample variance and kurtosis. A key contribution of this method is the ability to estimate these statistics without requiring access to an agent’s local mean, preserving a level of computational independence. We addressed both homogeneous systems and complex, multi-class scenarios where agents may share similar means but possess distinct variances or distribution shapes. By deriving the standard deviation of these estimators, we proposed an integrated method for constructing adaptive confidence intervals based on the sample mean, standard deviation, and kurtosis. This approach allows for effective graph-based collaboration in challenging cases previously unaddressed in the literature—specifically where data distributions differ only in higher-order statistics. To optimize the trade-off between convergence speed and the prevention of false pruning, we utilized weighted graphs with Gaussian kernels to reduce the influence of data points near interval boundaries. The theoretical framework was rigorously validated through numerical experiments, confirming the efficacy of using all three estimators in tandem for robust, collaborative learning.

\bibliographystyle{IEEEtran}
\bibliography{bibliography} 

\begin{IEEEbiographynophoto}{Nikola Stankovic}
 was born on December 28, 2001 in Podgorica, Montenegro. He received the B.Sc. degree in Computer Engineering and M.Sc. degree in Data Science and Engineering from Politecnico di Torino, Turin, Italy in 2024 and 2025, respectively. He is currently a PhD student at the University of Montenegro. His research interests include collaborative estimation, federated learning, artificial neural networks, and artificial intelligence approaches in engineering.
\end{IEEEbiographynophoto}

\end{document}